\def \Lop{L$_{5100}$}

\def \spitzer     {{\it Spitzer }}
\def\Msun{\ifmmode M_{\odot} \else $M_{\odot}$\fi}
\documentclass[oldversion]{aa}
\usepackage{graphicx}
\usepackage{longtable,lscape}
\usepackage{txfonts}
\usepackage{natbib}
\begin{document}
   \title{Dust covering factor, silicate emission and star formation in luminous QSOs}

   \author{R. Maiolino\inst{1}
	  \and
	  O. Shemmer\inst{2}
          \and
          M. Imanishi\inst{3}
	  \and
	  Hagai Netzer\inst{4}
	  \and
	  E. Oliva\inst{5}
	  \and
	  D. Lutz\inst{6}
	  \and
	  E. Sturm\inst{6}
}

   \offprints{R. Maiolino}

   \institute{INAF - Osservatorio Astronomico di Roma, via di Frascati 33,
    00040 Monte Porzio Catone, Italy 
         \and
	 Department of Astronomy and Astrophysics, 525 Davey Laboratory,
	 Pennsylvania State University, University Park, PA 16802
	 \and
	 National Astronomical Observatory, 2-21-1, Osawa, Mitaka,
	 Tokyo 181-8588, Japan
	 \and
	 School of Physics and Astronomy and the Wise Observatory ,
	 Tel-Aviv University, Tel-Aviv 69978, Israel
	 \and
	 INAF - Telescopio Nazionale Galileo, PO Box 565,
	 38700 Santa Cruz de La Palma, Tenerife, Spain
	 \and
     Max-Planck-Institut f\"{u}r Extraterrestrische Physik, D-85741 Garching,
	 Germany
             }

   \date{Received ; accepted }

 
   \abstract
    {We present \spitzer IRS low resolution, mid-IR spectra of
     a sample of 25 high luminosity QSOs at 2$<$z$<$3.5. When combined with
     archival IRS observations of local, low luminosity type-I active galactic nuclei (AGNs),
     the  sample spans five orders of magnitude in
     luminosity. 
  We find that the continuum dust thermal emission at
     $\rm \lambda _{rest}=6.7\mu m$ is correlated
     with the optical luminosity, following the non-linear relation
     $\rm \lambda L_{\lambda}(6.7\mu m) \propto \lambda
     L_{\lambda}(5100\AA)^{0.82}$.
     We also find an anti correlation between $\rm \lambda
     L_{\lambda}(6.7\mu m)/\lambda L_{\lambda}(5100\AA)$ and
	 the [O{\sc iii}]$\lambda$5007 line luminosity. These effects are
   interpreted as a decreasing covering factor of the circumnuclear
   dust as a function of luminosity. Such a result is in agreement
   with the decreasing fraction
   of absorbed AGNs as a function of luminosity recently found in
   various surveys. In particular, while X-ray surveys find a decreasing
   covering factor of the absorbing gas as a function of luminosity,
   our data provides an independent and
   complementary confirmation by finding a decreasing covering factor of dust.
   We clearly detect the silicate emission feature in
   the average spectrum, but also in four individual objects. These
   are the Silicate emission in the most luminous
   objects obtained so far. When combined with the silicate emission
   observed in local, low luminosity type-I AGNs, we find that the
   silicate emission strength 
   is correlated  with luminosity.
    The silicate strength of all type-I AGNs
   also follows  a positive
   correlation with the black hole mass and with the accretion rate. 
   The Polycyclic Aromatic Hydrocarbon (PAH) emission
   features, expected from starburst activity, are not detected in the
   average spectrum of luminous, high-z QSOs. The
    upper limit inferred
   from the average spectrum points to a ratio between PAH luminosity
   and QSO optical luminosity significantly lower than observed in
   lower luminosity AGNs, implying that the correlation between star
   formation rate and AGN power saturates at high
   luminosities.}

   \maketitle
%

\section{Introduction}
\label{sec_intro}

The mid-IR (MIR) spectrum of AGNs contains a wealth of information
which is crucial to the understanding of their inner
region. The observed prominent continuum emission is due to circumnuclear dust
heated to a temperature of several hundred degrees by the nuclear,
primary optical/UV/X-ray
source (primarily the central accretion disk); therefore, the MIR
continuum provides information on the amount and/or covering factor of the circumnuclear dust.
The MIR region is also rich of several emission features which are important
 tracers of the ISM. Among the dust features, the
Polycyclic-Aromatic-Hydrocarbon bands (PAH, whose most prominent feature is
at $\rm \sim 7.7 \mu m$) are emitted by very small carbon grains
excited in the Photo Dissociation Regions, that are tracers of
star forming activity \citep[although PAHs may not be
reliable SF tracers for compact HII regions
or heavily embedded starbursts,][]{peeters04,forster04}. 
Additional MIR dust features are the Silicate
bands at $\rm \sim 10 \mu m$ and at $\rm \sim 18 \mu m$, often seen in
absorption in obscured AGNs and in luminous IR galaxies.

Major steps forward in this field were achieved thanks to the \spitzer\
Space Observatory, and to its infrared spectrometer, IRS, which allows
a detailed investigation of the MIR spectral features in a
large number of sources. In particular, IRS
allowed the detection of MIR emission lines in several AGNs
\citep[e.g. ][]{armus04,haas05,sturm06a,weedman05},
the detection of PAHs in local PG QSOs
\citep{schweitzer06}, the first detection of the Silicate
feature in emission \citep{siebenmorgen05,hao05}, as well as detailed
studies of the silicate strength in various classes of sources
\citep{spoon07,hao07,shi06,imanishi07}.

However, most of the current \spitzer IRS studies have focused on local
and modest luminosity AGNs (including low luminosity QSOs),
with the exception of a few bright, lensed
objects at high redshift \citep{soifer04,teplitz06,lutz07}.
We have obtained
short IRS integrations of a sample of 25 luminous AGNs (hereafter QSOs) at high redshifts
with the goal of extending the investigation of the MIR properties
to the high luminosity range. The primary goals
were to investigate the covering factor of the circumnuclear dust
 and the dependence of the star formation rate (SFR), as traced by the
PAH features, on various quantities such as metallicity, narrow line
luminosity, accretion rate and black hole mass.
In combination with lower luminosity AGNs obtained by previous IRS
studies, our sample spans about 5 orders of magnitude in luminosity.
This allows us to look for the
dependence of the covering factor on luminosity and black
hole mass. We also search for the silicate emission and PAH-related properties although
the integration times  were too short, in most cases, to unveil the properties of
individual sources.

In Sect.~\ref{sec_obs} we discuss the sample selection, the observations and the
data reduction. In Sect.~\ref{sec_main_res} we describe the spectral analysis
and the main observational results, and in Sect.~\ref{sec_comp_lowl}
 we include additional data on local, low luminosity sources
from the literature and from the \spitzer\ archive.
The dust covering factor is discussed in Sect.~\ref{sec_dust_cov},
 the properties of the Silicate
emission feature in Sect.~\ref{sec_si} and the constraints on the star
formation in Sect.~\ref{sec_pah}.
The conclusion
are outlined in Sect.~\ref{sec_conc}.  Throughout the paper we assume a concordance
$\Lambda$-cosmology with $\rm H_0 = 71~km~s^{-1}~Mpc^{-1}$, $\rm
\Omega _m = 0.27$ and $\rm \Omega _{\Lambda} = 0.73$
\citep{spergel03}.

\section{Sample selection, observations and data reduction}
\label{sec_obs}

High redshift, high luminosity QSOs in our sample were mostly drawn
from \cite{shemmer04} and from \cite{netzer04}. The latter
papers presented near-IR spectra (optical rest-frame) of a large
sample of QSOs at 2$<$z$<$3.5, which were used to obtain detailed
information on the black hole (BH) mass (by means of the width of the
H$\beta$ line), on the accretion rate and on the strength of the
narrow emission line [OIII]$\lambda$5007. The sample contains also infrared data
on two sources from \cite{dietrich02} and a few additional QSOs in the
same redshift range, for which near-IR spectra where obtained after
\cite{shemmer04}, but unpublished yet. This sample
allows us not only to extend the investigation of the MIR properties
as a function of luminosity, but also to relate those properties to other
physical quantities such as BH mass, accretion rate and luminosity of
the narrow line region. In total our sample includes
25 sources which are listed in Table~\ref{tab_opt}.
Note that the QSOs in \cite{shemmer04} and in \cite{netzer04} were
extracted from optically or radio selected catalogs, without any
pre-selection in terms of mid- or far-IR brightness. Therefore, the
sample is not biased in terms of star formation or dust
content in the host galaxy.

We observed these QSOs with the Long-Low resolution module of the
\spitzer\ Infrared Spectrograph IRS \citep{houck04}, covering the
wavelength range 22--35$\mu$m, in staring mode. Objects were acquired
by a blind offset from a nearby, bright 2MASS star, whose location and
proper motion were known accurately from the Hipparcos catalog. We
adopted the ``high accuracy'' acquisition procedure, which provides a
slit centering good enough to deliver a flux calibration accuracy
better than 5\%. The integration time was of 12~minutes on
source, with the exception of seven which were observed only 4
minutes each\footnote{More specifically: LBQS0109+0213, $[$HB89$]$1318-113,
$[$HB89$]$1346-036,SBS1425+606,$[$HB89$]$2126-158,
2QZJ222006.7-280324,VV0017.}

We started our reduction from the Basic Calibrated Data (BCD).  For
each observation, we combined all images with the same position on the
slit.
Then the sky background was subtracted by using pairs of frames where
the sources appears at two different positions along the slit.
The spectra were cleaned for bad, hot and rogue pixels by using the
IRSCLEAN algorithm. The monodimensional spectra were then extracted by means of the
SPICE software.

\section{Analysis}
\label{sec_analysis}

\subsection{Main observational results}
\label{sec_main_res}

All of the objects were clearly detected. In Tab.~\ref{tab_ir} we
list the observed continuum flux densities at the observed wavelength
corresponding to $\rm
\lambda _{rest} = 6.7 \mu m$.  This wavelength was chosen both because it is
directly observed in the spectra of all objects
and because it is far from the Silicate feature
and in-between PAH features. Thus the determination of L(6.7$\mu$m) should be
little affected by uncertainties in the subtraction of the
starburst component (see below).
For two of the radio-loud objects ($[$HB89$]$0123+257 and TON618) the MIR flux
lies on the extrapolation of the synchrotron radio emission and
therefore the former is also probably non-thermal. Since in this paper
we are mostly interested in the thermal emission by dust, the latter
two objects are not used in the statistical analysis. For the other
two radio loud QSOs, the extrapolation of the radio spectrum falls
below the observed MIR emission and the latter is little
affected by synchrotron contamination.

   \begin{figure}
   \centering
   \includegraphics[bb=30 150 485 660, width=9cm]{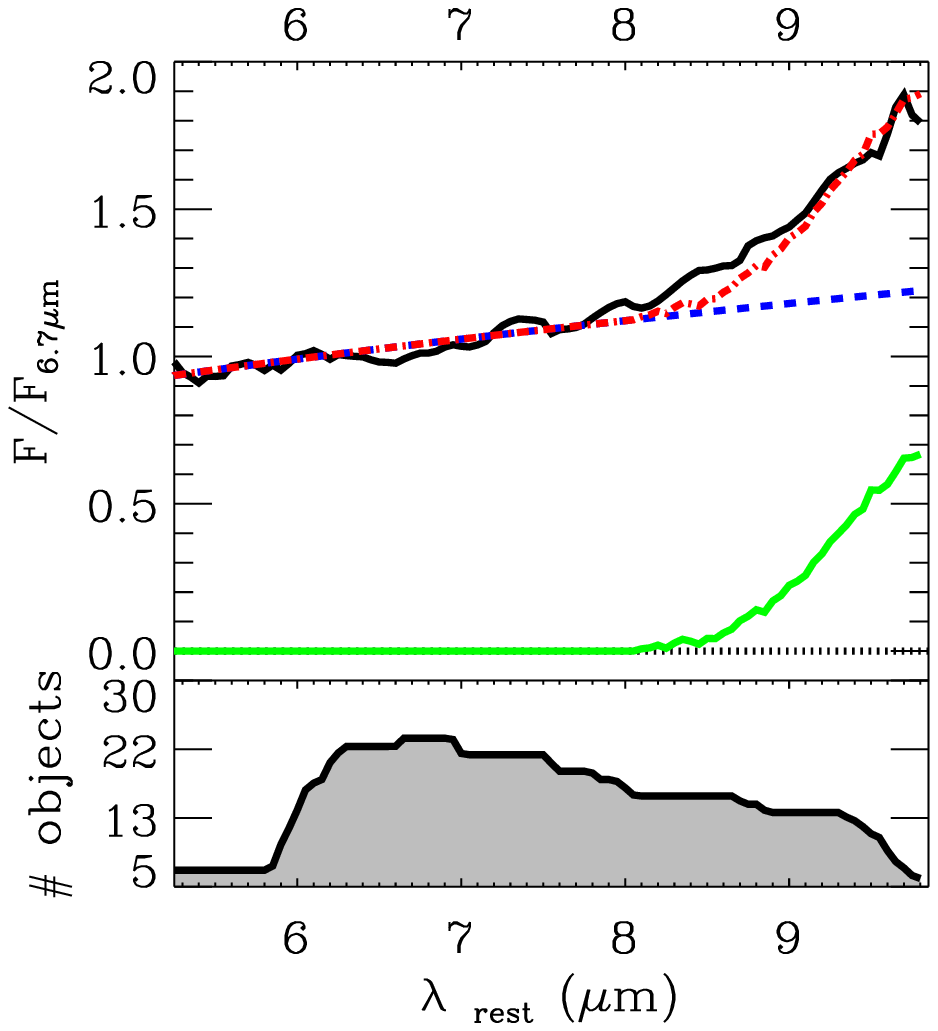}
   \caption{Average spectrum of all high-z, luminous QSOs in our
     sample, normalized to the flux at 6.7$\mu$m (black solid
     line). The blue dashed line indicates the power-law fitted to
     the data at $\rm \lambda < 8 \mu m$; the green solid line is
     the fitted silicate emission and the red, dot-dashed line is the
     resulting fit to the stacked spectrum (sum of the power-law and
     silicate emission). The bottom panel indicates the number of
     objects contributing to the stacked spectrum at each wavelength.}
         \label{fig_av}
   \end{figure}

   \begin{figure*}[!]
   \centering
   \includegraphics[bb=20 170 540 680,width=17cm]{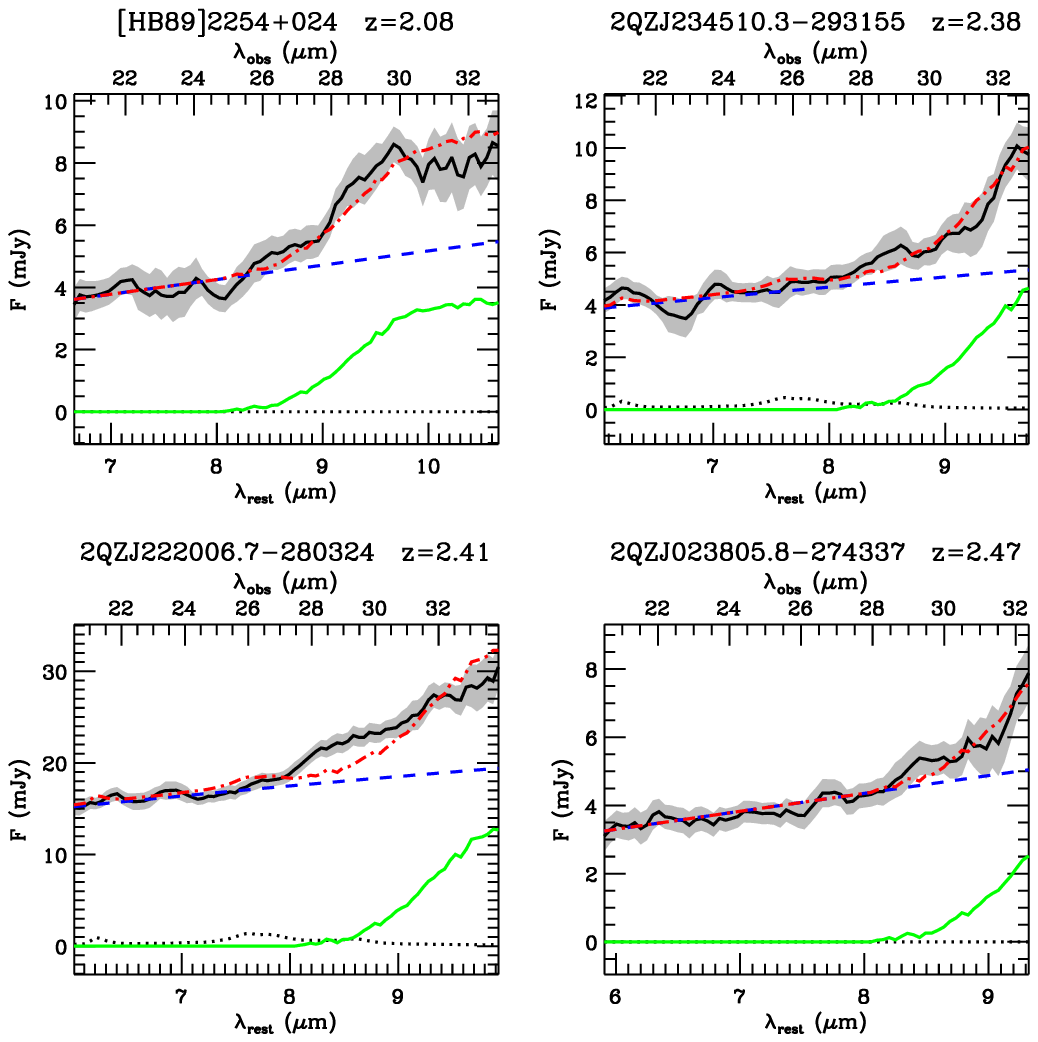}
   \caption{IRS spectra of four individual high-z luminous QSOs
     showing evidence for silicate emission. The black
     solid lines indicate the IRS spectra smoothed with a 5 pixels
     boxcar. The shaded areas indicate the flux uncertainty.
     The blue dashed line and the green solid line are the power-law
     and the silicate emission components of the fits.
     The black dotted line shows the starburst component,
     which is formally required by the fit, but statistically not
     significant. The red dot-dashed lines are the global
      fits to the observed spectra.}
       \label{fig_4si}
    \end{figure*}


   Fig.~\ref{fig_av} shows the mean spectrum of all sources in the
   sample, except for the two which are likely dominated by synchrotron emission. Each spectrum
   has been normalized to 6.7$\mu$m prior to averaging. The bottom
   panel shows the number of sources contributing to the mean spectrum
   in different spectral regions. We only consider the rest frame
   spectral range where at least 5 objects contribute to the mean
   spectrum. The spectrum at $\rm \lambda < 8 \mu m$ has been fitted with a simple power-law.
   While other workers in
   this field assumed more complicated  continuum (e.g. spline,
   polynomial) we do not consider it justified  given the limited wavelength range
   of our spectra.
    The extrapolation of the continuum to
   10$\mu$m clearly reveals an excess identified with  Silicate
   emission.  Fitting and
   measuring the strength of this feature is not easy given
   the limited rest-frame spectral coverage of our spectra.
   Therefore, we resort to the use of templates. In
   particular, we fit the Silicate feature by using as a template the
   (continuum-subtracted) silicate feature observed in the average
   spectrum of local QSOs as obtained by the QUEST project
   \citep{schweitzer06} and kindly provided by M. Schweitzer.
    The template Silicate spectrum, with the best fitting
   scaling factor is shown in green in Fig.~\ref{fig_av}, while the
   red dot-dashed line shows the resulting fit including the
   power-law.  We adopt the definition of ``silicate strength'' given
   in \citep{shi06} which is the ratio between the maximum of the
   silicate feature and the interpolated featureless continuum at the
   same wavelength. In the QSO-QUEST template the maximum of the
   Silicate feature is at $\rm 10.5\mu m$. This wavelength is
   slightly outside the band covered by our spectra but the uncertainty on the
   extrapolation is not large (the latter is included in the error estimate
   of the silicate strength).  The silicate strength in the
   mean spectrum is 0.58$\pm$0.10 (Tab.~\ref{tab_ir}).

   We note that the average spectrum does
   not show evidence for PAH features at 7.7$\mu$m and 6.2$\mu$m. Such features are
    observed in lower luminosity AGNs.
    More specifically, a starburst template \citep{sturm00} is
   not required by the fit shown in Fig.~\ref{fig_av}. In Sect.~\ref{sec_pah}
   we will infer an
   upper limit on the PAH luminosity and discuss its implication.

   We clearly detect the blue wing of the
   silicate feature in four individual spectra,
   which are shown in Fig.~\ref{fig_4si}. These spectra were fitted
   with a power-law and a silicate template exactly as the stacked
   spectrum. The resulting values for the Silicate strength are given
   in Tab.~\ref{tab_ir}.  The presence
   of  silicate emission in all other cases is poorly constrained (or totally
   unconstrained) either because of low signal-to-noise (S/N) or because of
   a lack of spectral coverage. The one exception is Ton\,618 which has a high
   S/N spectrum and a redshift (z=2.22) appropriate to
   observe the Silicate 10.5 $\mu$m feature. No silicate emission is detected in this case,
   but note that this is not expected since the MIR radiation of this source
   is probably dominated by synchrotron emission.

   Tables~\ref{tab_opt} and \ref{tab_ir} list
   the more important MIR  information on the sources
   and physical properties deduced from
   the rest-frame optical spectra and obtained from \cite{shemmer04}:
   optical continuum luminosity $\rm \lambda L_{\lambda}(5100\AA )$,
   [OIII]$\lambda$5007 line luminosity, BH mass and Eddington
   accretion rate $\rm L/L_{Edd}$.

\subsection{A comparison with MIR properties of lower luminosity AGNs}
\label{sec_comp_lowl}

To compare the MIR properties of our luminous QSOs with those of lower
luminosity sources we have included in our study the IRS/MIR spectra
of various low redshift, lower luminosity type-I AGNs.
We purposely avoid type-II sources because of the additional complication due to
absorption along the line of sight.

We have used data  from \cite{shi06} who
analyze the intensity of the silicate features in several, local AGNs with
luminosities ranging from those of nearby Seyfert 1s  to intermediate luminosity QSOs.
 We discarded BAL QSOs (which are known to have intervening gas and dust
absorption) as well as dust reddened type-I nuclei (e.g. 2MASS red
QSOs). We also discard those cases (e.g. 3C273) where the optical and
MIR continuum is likely dominated by synchrotron radiation. Note that
\cite{shi06} selected type-I objects with ``high
brightness'' and, therefore, low-luminosity AGNs tend to be excluded
from their sample.

\cite{shi06} provide a measure of the silicate feature strength
(whose definition was adopted also by us).
The continuum emission at 6.7$\mu$m
was measured by us from the archival spectra.  We also subtracted from
the 6.7$\mu$m emission the possible contribution of a starburst
component by using the M82 template.
We estimate the host galaxy contribution (stellar photospheres)
in all sources to be negligible.

We include in the sample of local Sy1s also some IRS spectra taken from
the sample of \cite{buchanan06}, whose spectral parameters were determined
by us from the archival spectral, in the same manner as for the \cite{shi06}
spectra. As for the previous sample, we discarded
reddened/absorbed sources as well as those affected by synchrotron emission.
As discussed in \cite{buchanan06}, these spectra are affected by significant
flux calibration uncertainties, due to the adopted mapping technique. Therefore,
the spectra were re-calibrated by using IRAC photometric images. We discarded
objects for which IRAC data are not available or not usable (e.g. because saturated).
Finally, we also discarded data for which optical spectroscopic data are not
available (see below).

The mid-IR parameters of the sources in both samples
are listed in Tab.~\ref{tab_ir}.

Optical data were mostly taken from \cite{marziani03} and BH masses
and Eddington accretion rates inferred as in \cite{shemmer04}.
The resulting parameters are listed in Table~\ref{tab_opt}.

The type-I sources in \cite{shi06} and \cite{buchanan06} are only used for the
investigation of the covering factor and silicate strength, which are
the main aims of our work. The  \cite{shi06} and \cite{buchanan06} samples are
not suitable
for investigating the PAH features because most of
these objects are at  small distances and the IRS slit misses most
of the star formation regions in the host galaxy. For what concerns
the the PAH emission, we use the data in
\cite{schweitzer06} who performed a detailed analysis of the PAH
features in their local QSOs sample. The slit losses in those sources are
minor. The \cite{schweitzer06} sample is also used for the investigation of the
MIR-to-optical luminosity ratio. The mid-IR data of this sample are not listed
in Tab.~\ref{tab_ir}, since such data are
already reported in \cite{schweitzer06} and in \cite{netzer07}.


\section{Discussion}
\label{sec_disc}

\subsection{Dust covering factor}
\label{sec_dust_cov}

\subsubsection{Covering factor as a function of source luminosity and BH mass}
\label{sec_dust_cov_L}

The main assumption used here is that the
covering factor of the circumnuclear dust is given by the
ratio of the thermal infrared emission
to the primary AGN radiation. The latter is mostly the ``big blue bump'' radiation
with additional contribution from the optical
and X-ray wavelength ranges \citep{netzer90}.

Determining the integral of the AGN-heated dust emission, and
disentangling it from other spectral components is not simple.  The
FIR emission in type-I AGNs is generally dominated by a starburst component, even
in QSOs \citep{schweitzer06}. In lower luminosity AGNs,
 the near-IR emission may be affected by  stellar
emission in the host galaxy, while in QSOs the near-IR light is often
contributed also by the direct primary radiation.
The MIR range ($\rm \sim 4-10\mu m$) is where the contrast between AGN-heated
dust emission and other components is maximal.
This spectral region contains various spectral features, like PAHs
and silicate emission, yet MIR spectra allow us to disentangle and
remove these  components, and determine the hot dust continuum.
In particular, by focusing on the continuum emission at
6.7$\mu$m, the uncertainty in the removal of PAH emission is minimized,
while the contribution from the Silicate emission is totally
negligible at this wavelength (note that such a spectral decomposition
is unfeasible with photometric data).
 If the spectral shape of the
AGN-heated dust does not change from object to object (and in
particular it does not change significantly with luminosity), then the
6.7$\mu$m emission is a proxy of the global circumnuclear hot dust
emission. It is possible to infer a quantitative relation between
L(6.7$\mu$m) and the total AGN-heated hot dust emission through the work of
\cite{silva04}, who use observations of
 various nearby AGNs to determine their average, {\it nuclear} IR SED
(divided into absorption classes).
From their type I AGNs SED, we find that the integrated nuclear, thermal IR bump
is about $\rm \sim 2.7~\lambda L_{\lambda}(6.7\mu m)$. This ratio is also consistent with
that found in the QUEST QSO sample, once the contribution by the silicate features is subtracted.

Regarding the primary optical-UV radiation, determining its
integrated flux would require observations of the entire intrinsic spectral energy distribution (SED)
from the far-UV to the
near-IR. This is not available for most sources in our sample. Moreover,
there are indications that the SED is luminosity dependent
\citep[e.g.][]{scott04,shang05}
and thus an estimate of the bolometric luminosity of the primary continuum
based on the observed  luminosity in a certain band is somewhat uncertain.
Notwithstanding these limitations, we assume in this work that
the optical continuum luminosity can be
used as a proxy of the bolometric luminosity of the primary continuum.
We use the
continuum luminosity at the rest frame wavelength of 5100\AA, ($\lambda L_{\lambda}$, hereafter \Lop)
because it is directly measured for all sources in our sample.
We can infer the ratio between
the bolometric luminosity and \Lop\ from the mean
spectrum of QSOs obtained in recent studies, mostly the results of \cite{scott04} and
\cite{richards06}. These studies indicate a bolometric correction in the
range of 5--9. Here we chose, rather arbitrarily, the mean value of 7.

   \begin{figure}[!]
   \centering
   \includegraphics[bb=30 160 500 610,width=9cm]{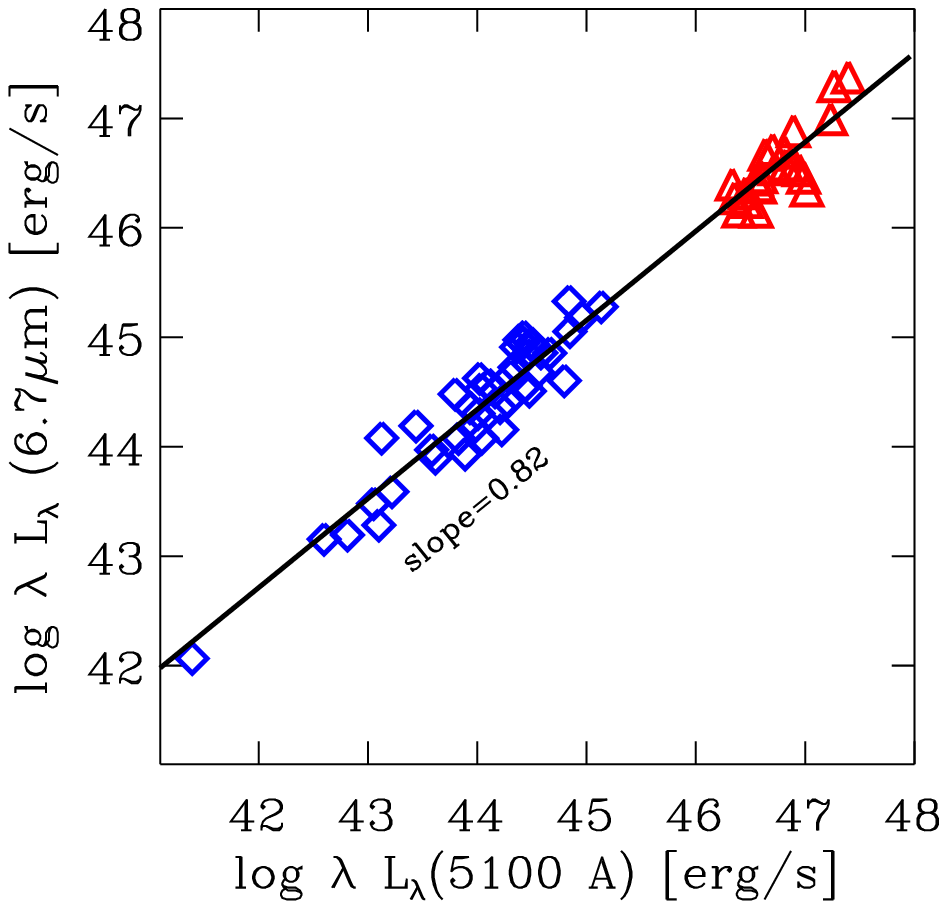}
   \caption{MIR continuum luminosity at 6.7$\mu$m versus optical
     continuum luminosity at 5100\AA.  Red triangles mark high-z,
     luminous QSOs and blue diamonds mark low redshift type-I AGNs.
      The black solid line is
     a fit to the data, which has a slope with a power index of 0.82.}
         \label{fig_miropt}
   \end{figure}

   \begin{figure*}[!]
   \centering
   \includegraphics[bb=35 195 540 630,width=19cm]{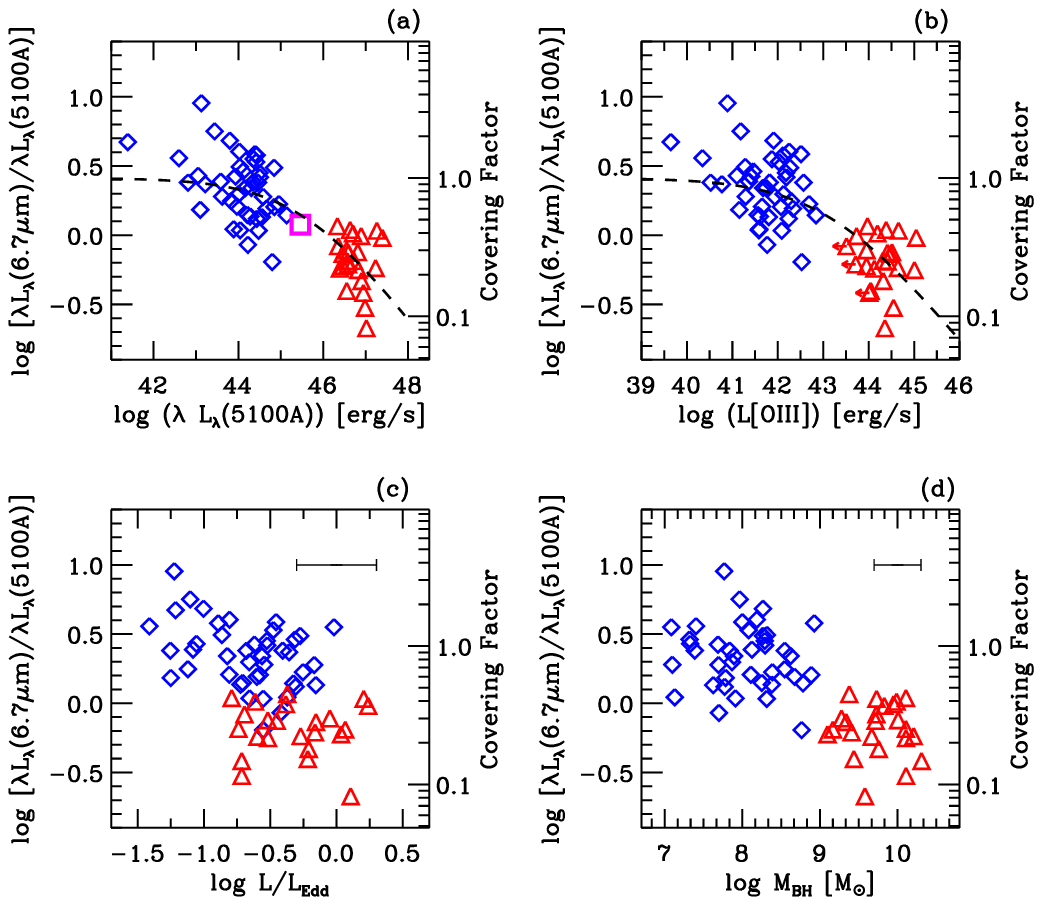}
   \caption{MIR--to--optical continuum luminosity $\rm [\lambda
     L_{\lambda}(6.7\mu m)]/ [\lambda L_{\lambda}(5100\AA )]$ versus
     (a) continuum optical luminosity, (b) [OIII]$\lambda$5007 line
     luminosity, (c) accretion rate $\rm L/L_{Edd}$ and (d) black hole
     mass.  Symbols are the same as in Fig.~\ref{fig_miropt}. The
     magenta square indicates the location of the mean SDSS QSO SED in
     \cite{richards06}. The horizontal error bars in panels (c) and (d) indicate
	 conservative uncertainties on the accretion rates and BH masses.
	 The right hand side axis on each panel shows
     the circumnuclear dust covering factor inferred from Eq.~\ref{eq_cf}.  The
     dashed lines in panels (a) and (b) shows the fit resulting from
     the analytical forms in Eqs.~\ref{eq_f2_opt} and \ref{eq_f2_oiii}, respectively.}
         \label{fig_cf}
   \end{figure*}

The comparison of  L(6.7$\mu$m) and \Lop\ is our way of deducing the
 hot dust covering factor.
Fig.~\ref{fig_miropt} shows the $\rm \lambda L_{\lambda}(6.7\mu m)$
versus \Lop\ for our sample.  Red triangles are
the high-z, luminous QSOs and blue diamonds are local type-I AGNs.
 Not surprisingly, the
two quantities show a good correlation. However, the very large luminosity
range spanned by our sample allows us to clearly state that the
correlation is not linear, but has a slope
 $\rm \alpha = 0.82\pm 0.02$ defined by
\begin{equation}
  \rm log[\lambda L_{\lambda}(6.7\mu m)] = K +
\alpha ~log[\lambda L_{\lambda}(5100\AA )]
\label{eq_miropt}
\end{equation}
 where $\rm K = 8.36\pm 0.80$ and  luminosities are expressed in $\rm erg ~s^{-1}$. This
indicating that the MIR, reprocessed emission increases more slowly
than the  primary luminosity.

The same phenomenon is observed in a cleaner way in Fig.~\ref{fig_cf}a,
where the ratio between the two continuum luminosities is plotted as a
function of \Lop. There is a clear anti-correlation
between the MIR--to--optical ratio and optical luminosity.  Fig.~\ref{fig_cf}b
shows the same MIR--to--optical ratio as a function of
L([OIII]$\lambda$5007) as an alternative tracer of the global
AGN luminosity \citep[although the latter is not a linear tracer of
the nuclear luminosity, as discussed in ][]{netzer06}, which displays
the same anti-correlation as for the continuum optical
flux. Spearman-rank coefficients and probabilities for these
correlations are given in Tab.~\ref{tab_corr}.

According to the above discussion, an obvious
interpretation of the decreasing MIR--to--optical ratio is that the
covering factor of the dust surrounding the AGN decreases with
luminosity. In particular, if the covering factor is proportional to
the MIR--to--optical ratio, then Figs.~\ref{fig_cf}a-b indicate that
the dust covering factor decreases by about a factor 10 over the
luminosity range probed by us.

It is possible to convert the
MIR--to--optical luminosity ratio into absolute dust covering factor
by assuming ratios of broad band to monochromatic continuum
luminosities observed in AGNs, as discussed above. In particular, by using the
6.7$\mu$m--to--MIR and 5100\AA--to--bolometric luminosity ratios reported above,
we obtain that the absolute value of
the dust covering factor (CF) can be written as:
\begin{equation}
\rm CF(dust) \approx 0.39 \cdot
\frac{\lambda L_{\lambda}(6.7\mu m)}{\lambda L_{\lambda}(5100\AA )} ~.
\label{eq_cf}
\end{equation}

In Fig.~\ref{fig_cf} the axes on the right hand side provide the dust
covering factor inferred from the the equation above. A fraction of objects
have covering factor formally larger than one, these could be
due to uncertainties in the observational data, or nuclear SED
differing from the ones assumed above, or to optical variability, as
discussed in Sect.~\ref{sec_model_unc}.  The dust covering factor ranges from about
unity in low luminosity AGNs
to about 10\% in high luminosity QSOs. As it will be discussed in
detail in Sect.~\ref{sec_comp_prev},
the dust covering factor is expected to be equal to the fraction of type 2
(obscured) AGNs relative to the total AGN population. The finding of a large 
covering factor in low luminosity AGNs is in agreement
with the large fraction of type 2 nuclei observed in local Seyferts
\citep[$\sim 0.8$,][]{maiolino95}.

A similar result has been obtained, in an independent way, through the
finding of a systemic decrease of the the obscured to unobscured AGN
ratio as a function of luminosity in various surveys. The comparison
with these results will be discussed in more detail in the next
section.

The physical origin of the decreasing covering factor is still
unknown. One
possibility is that higher luminosities imply a larger dust
sublimation radius: if the obscuring medium is distributed in a disk
with constant height, then a larger dust sublimation radius would
automatically give a lower covering factor of dust at higher
luminosities \citep{lawrence91}.  However, this effect can only
explain the decreasing covering factor with luminosity for the dusty
medium, and not for the gaseous X-ray absorbing medium. Moreover,
\cite{simpson05} showed that the simple scenario of such a ``receding
torus'' is unable to account for the observed dependence of the type 2
to type 1 AGN ratio as a function of luminosity. 

Another possibility
is that the radiation pressure on dust is stronger, relative to the BH
gravitational potential, in luminous AGNs
\citep[e.g. ][]{laor93,scoville95}, thus sweeping away circumnuclear
dust more effectively.  In this scenario a more direct relation of the
covering factor should be with $\rm L/L_{Edd}$, rather than with
luminosity. Our sample does not show  such a relation, as illustrated in
Fig.~\ref{fig_cf}c. However, the uncertainties on the accretion
rates (horizontal error bar in Fig.~\ref{fig_cf}c) may hamper the identification of
such a correlation.

\cite{lamastra06} proposed that,
independently of luminosity, the BH gravitational potential is
responsible for flattening the circumnuclear medium, so that larger BH
masses effectively produce a lower covering factor. According to this
scenario, the relation between covering factor and luminosity is only
an indirect one, in the sense that more luminous AGN tend to have
larger BH masses (if the Eddington accretion rate does not change
strongly on average).
Fig.~\ref{fig_cf}d shows the MIR--to--optical ratio (and dust covering
factor) as a function of BH mass, indicating a clear
(anti-)correlation between these two quantities.  However, the
correlation is not any tighter than the relation with luminosity in
Fig.~\ref{fig_cf}a-b, as quantified by the comparison of the
Spearman-rank coefficients and probabilities in Tab.~\ref{tab_corr}.
The degeneracy between luminosity and BH mass prevents us to
discriminate which of the two is the physical quantity driving the
relation.

\subsubsection{Model uncertainties}
\label{sec_model_unc}

In this section we discuss some possible caveats in our
interpretation of the MIR--to--optical ratio as an indicator of the hot dust
covering factor.

   \begin{figure}[!]
   \centering
   \includegraphics[bb=15 162 500 625,width=7.5cm]{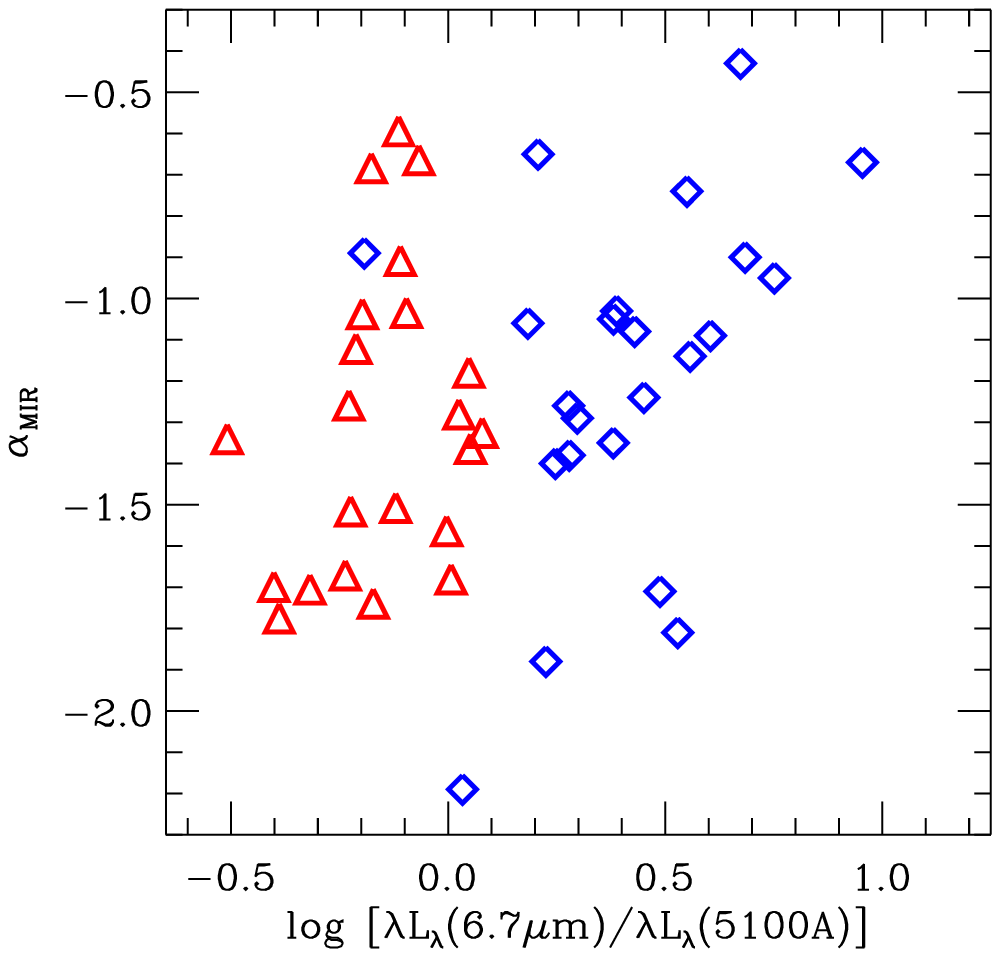}
   \caption{Mid-IR spectral slope (5--8$\mu$m) versus
   $\rm \lambda L_{\lambda}(6.7\mu m)/\lambda L_{\lambda}(5100\AA)$ ratio. Symbols
   are as in Fig.~\ref{fig_miropt}. No clear correlation is observed between these two
   quantities (see also Tab.~\ref{tab_corr}). Note that the exceptional object
   SDSSJ173352.22+540030.5 is out of scale, and it is discussed in the text.}
         \label{fig_alphair}
   \end{figure}

   \begin{figure}[!]
   \centering
   \includegraphics[bb=15 162 500 625,width=7.5cm]{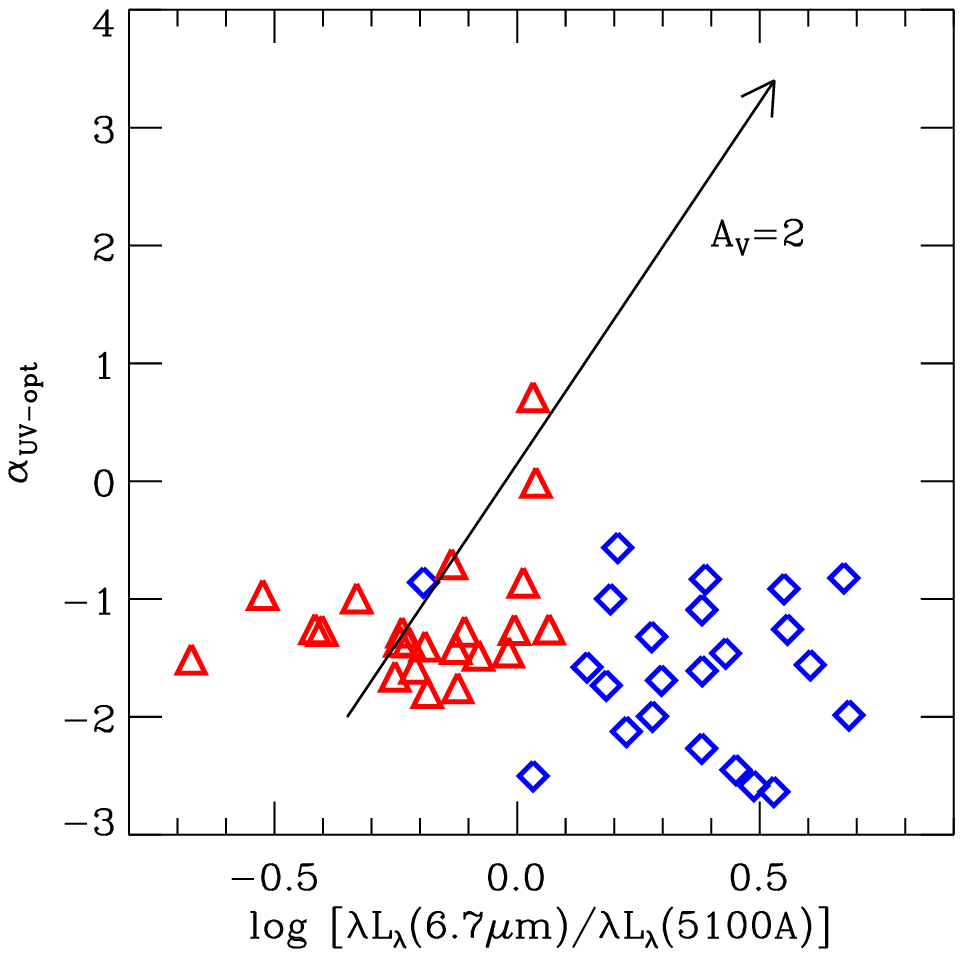}
   \caption{Optical-to-UV spectral slope (1450--5100\AA) versus
   $\rm \lambda L_{\lambda}(6.7\mu m)/\lambda L_{\lambda}(5100\AA)$ ratio. Symbols
   are as in Fig.~\ref{fig_miropt}. The arrow indicates the effect of dust reddening
   with $\rm A_V=2$~mag. The data do not show any evidence for dust
   reddening effects.}
         \label{fig_alphauv}
   \end{figure}

Our analysis assumes that the shape of the hot dust IR spectrum 
SED is not luminosity dependent.
  However, an alternative interpretation of the trends
observed in Fig.~\ref{fig_cf} could be that the dust
temperature distribution changes with luminosity. Yet, to explain the decreasing
6.7$\mu$m to 5100\AA \ flux ratio in terms of dust temperature would
require that the circumnuclear dust is cooler at higher luminosities.
This, besides being contrary to
expectations, is ruled out by the observations which show no clear
correlation between the mid-IR continuum slope (which is associated with
the average dust temperature) and the MIR--to--optical ratio, as illustrated
in Fig.~\ref{fig_alphair} and in Tab.~\ref{tab_corr}.
The one remarkable exception is SDSSJ173352.22+540030.5, which has the most negative
mid-IR continuum slope ($\alpha _{MIR}= -2.88$, i.e. an inverted spectrum
in $\rm \lambda L_{\lambda}$)
and the lowest MIR-to-optical ratio
of the whole sample ($\rm \lambda L_{\lambda}(6.7\mu m)/\lambda L_{\lambda}(5100\AA) =
0.22$), which is out of scale in Fig.~\ref{fig_alphair}. This QSO may be totally
devoid of circumnuclear hot dust, and its MIR emission may simply be the continuation
of the optical ``blue-bump''. Similar high-z QSOs, with an exceptional deficiency
of mir-MIR flux, have been reported by \cite{jiang06}.

As explained earlier, there are indications that the UV-optical SED, and
hence the bolometric correction based to the observed \Lop, are luminosity dependent.
If correct, this would mean a smaller bolometric correction for higher
\Lop\ sources which would flatten the relationship found here
(i.e. will result in a slower decrease of the covering factor with
increasing \Lop). However, the expected range (a factor of at most 2
in bolometric correction) is much smaller than the deduced change
in covering factor.

Variability is an additional potential caveat because of the time delay
between the original \Lop\ ``input'' and the response of the dusty absorbing ``torus''.
While we do not have the observations to test this effect
(multi-epoch, high-quality
optical spectroscopic data are available only for a few sources in our sample),
we note that
the location of the 6.7$\mu$m emitting gas from the central accretion disk is at least
several light years and thus L(MIR) used here reflects the mean \Lop\
in most sources.
We expect that the average luminosity of a large sample
will not be affected much by individual source variations. Moreover, in the
high luminosity sources of our sample we do not expect much variability
(since luminous QSOs are known to show little or no variability).

An alternative, possible explanation of the variation of the optical-to-MIR
luminosity ratio could be dust extinction affecting the observed optical flux.
Optical dust absorption increasing towards low luminosities may in principle explain the
trends observed in Fig.~\ref{fig_cf}. However, we have pre-selected the sample of local
QSOs and Sy1s to avoid objects showing any indication of absorption, thus probably
shielding us from such spurious
effects. Yet, we have further investigated the extinction
scenario by analyzing the optical-to-UV continuum shape of our sample. The
optical-UV continuum slope does not necessarily trace dust reddening, since intrinsic
variations of the continuum shape are known to occur, as discussed above. Variability
introduce additional uncertainties, since optical and UV data are not simultaneous.
However, if the variations of
$\rm \lambda L_{\lambda}(6.7\mu m)/\lambda L_{\lambda}(5100\AA)$ are mostly due
to dust reddening, one would expect the MIR-to-optical ratio to correlate with the
optical-UV slope, at least on average. We have compiled UV rest-frame
continuum fluxes (at $\rm \lambda _{rest}\sim 1450\AA $) from spectra in the literature
or in the HST archive. By combining such data with the continuum luminosities at 
$\rm \lambda _{rest}=5100\AA$ we inferred the optical-UV continuum
slope $\alpha_{opt-UV}$ defined as\footnote{Our definition of power law
index is linked to the $\alpha _{\nu}$ given in \cite{vandenberk01} by the relation
$\rm \alpha_{opt-UV} = -(\alpha _{\nu}+2)$. Our distribution of $\rm \alpha_{opt-UV}$
is roughly consistent (within the uncertainties and the scatter) with $\rm \alpha
_{\nu}=-0.44$ obtained by \cite{vandenberk01} for the SDSS QSO composite spectrum
\citep[see also ][]{shemmer04}.}
$\rm L_{\lambda}\propto \lambda ^{\alpha_{opt-UV}}$,
as listed in Tab.~\ref{tab_opt}. Fig.~\ref{fig_alphauv} shows $\rm \alpha_{opt-UV}$
versus $\rm \lambda L_{\lambda}(6.7\mu m)/\lambda L_{\lambda}(5100\AA)$. The
arrow indicates the effect of dust reddening with $\rm A_V=2$~mag \citep[by assuming a SMC
extinction curve, as appropriate for type 1 AGNs,][]{hopkins04}, which would be
required to account for the observed variations of
$\rm \lambda L_{\lambda}(6.7\mu m)/\lambda L_{\lambda}(5100\AA)$.
Fig.~\ref{fig_alphauv} does not show evidence for any (positive) correlation between
$\rm \alpha_{opt-UV}$
and $\rm \lambda L_{\lambda}(6.7\mu m)/\lambda L_{\lambda}(5100\AA)$
(see also Tab~\ref{tab_corr}). In particular, if the variation of MIR-to-optical ratio
(spanning more than a factor of ten) was due to dust reddening, we would expect
to find $\Delta \alpha_{opt-UV} > 5$ (as indicated by the arrow in
Fig.~\ref{fig_alphauv}), which is clearly not observed.
If any, the data show a marginal anti-correlation between $\rm \alpha_{opt-UV}$
and $\rm \lambda L_{\lambda}(6.7\mu m)/\lambda L_{\lambda}(5100\AA)$ (Tab.~\ref{tab_corr}),
i.e. opposite to that expected from dust reddening.

\begin{figure*}[!]
   \centering
   \includegraphics[bb=20 280 530 555,width=18cm]{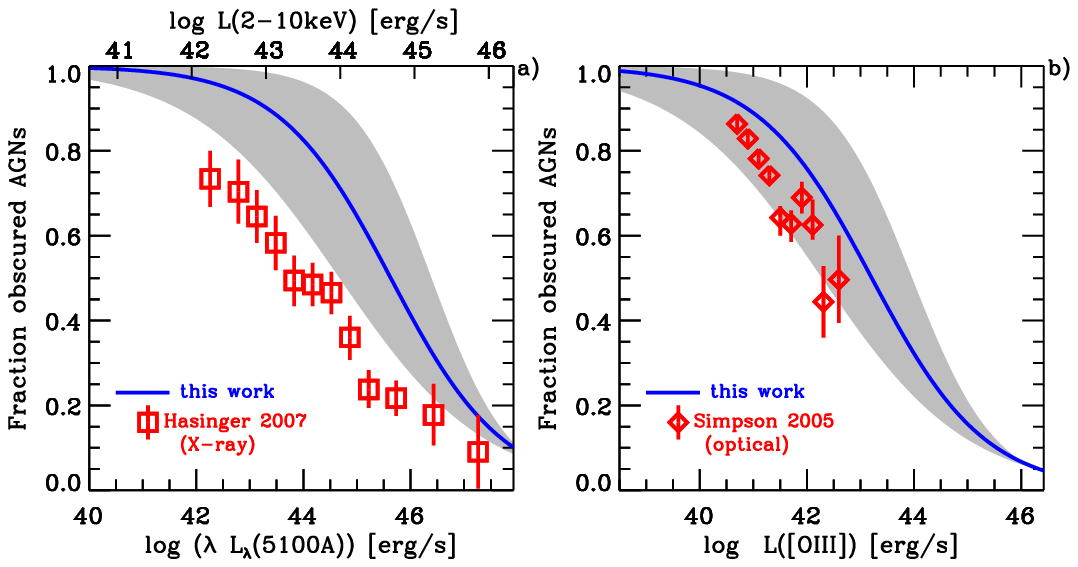}
   \caption{a) Fraction of obscured AGNs (relative to total) as a
     function of optical continuum luminosity. The blue line shows the
     fraction of obscured AGNs inferred from the hot dust covering
     factor with the analytical form of Eq.~\ref{eq_f2_opt}. The shaded
	 area is the uncertainty resulting from the plausible range of the
	 bolometric correction (see text). Points with error bars
     are the fraction of X-ray obscured AGNs as a function of X-ray
     luminosity inferred by Hasinger (2007).  The upper scale of the
     diagram gives the intrinsic hard X-ray luminosity; the
     (non-linear) correspondence between X-ray and optical luminosity
     is obtained from Eq.~\ref{eq_conv_ox}. b)
     Fraction of obscured AGNs as a function of L([OIII]$\lambda$5007).
     The blue line is the fraction of obscured AGNs
     inferred from the hot dust covering factor with the analytical
     form of Eq.~\ref{eq_f2_oiii}. Points with error bars show the fraction of type 2
     AGNs as a function of L([OIII]$\lambda$5007) inferred by
     \cite{simpson05}.  }
         \label{fig_fr_typ2}
   \end{figure*}

Finally, possible evolutionary effects on the dust covering factor
have not been considered.  We have been comparing local objects with
QSOs at z$\sim$2--3 yet assumed that only luminosity or BH
mass plays a role. \cite{lafranca05} and \cite{akylas06} find evidence
for an increasing fraction of obscured AGNs as a function of
redshift, a result which is still debated
 \citep[see ][]{ueda03,gilli07}. If the absorbing medium covering
 factor really increases with redshift,
 then the actual dependence of the
covering factor on luminosity would be stronger than shown
in Fig.~\ref{fig_cf}. Indeed, according to \cite{lafranca05} and
\cite{akylas06}, putative low-z counterparts of our high-z QSOs (matching
the same luminosities) should be affected by an even lower covering factor.
As a consequence, the diagrams in Figs.~\ref{fig_cf} and \ref{fig_fr_typ2}
should have even steeper trends once the data are corrected for such
putative evolutionary effects, thus strengthening our conclusions.

\subsubsection{Comparison with previous works}
\label{sec_comp_prev}

\cite{wang05} used IRAS and ISO mid-IR data of (mostly local) PG quasars 
to infer their dust covering factor. They find that the covering factor 
decreases as a function of X-ray luminosity. They probe a narrower
luminosity range with respect to our work,
nonetheless their results are generally consistent with ours,
although with significant scatter.

More recently \cite{richards06} derived the broad band SED of a
large sample of SDSS QSOs by including \spitzer\ {\it photometric} data.
Although
such data do not have spectroscopic information, they can be used to obtain a
rough indication of the dust covering factor in their QSO sample, to be compared
with our result. The average optical luminosity of the
\cite{richards06} sample is $\rm \langle log [\lambda
L_{\lambda}(5100\AA )]\rangle \sim 45.5~erg~s^{-1}$.
  From their mean
SED we derive $\rm \lambda L_{\lambda}(6.7\mu m)/\lambda
L_{\lambda}(5100\AA )=1.17$. The corresponding location on the diagram
of Fig.~\ref{fig_cf}a is marked with a magenta square, and it is in
agreement with the general trend of our data.

In a companion paper, \cite{gallagher07} use the same
set of data to investigate the MIR-to-optical properties as a function of luminosity.
They find a result similar to ours, i.e. the MIR-to-optical ratio decreases with
luminosity. However, they interpret such a result as a consequence of dust reddening
in the optical, since the effect is stronger in QSOs with redder optical slope.
As discussed in the previous section, our data do no support this scenario, at least
for our sample. In particular, the analysis of the optical-UV slope indicates that dust
reddening does not play a significant role in the variations of the
MIR-to-optical luminosity ratio. The discrepancy between our and \cite{gallagher07}
results may have various explanations.
The QSOs in the \cite{gallagher07} sample
span about two orders of magnitudes in luminosity, while
we have seen that to properly quantify the effect a wider luminosity
range is required. Moreover, the majority of their sources are clustered
around the mean luminosity of $\rm 10^{45.5}~erg~s^{-1}$. In addition, the
the lack of spectroscopic information makes it difficult to allow for the presence
of other spectral features such as the silicate emission which, as we show later,
is luminosity dependent. The lack of spectroscopic information
may be an issue specially for samples spanning a wide redshift range
\citep[as in][]{gallagher07},
where the photometric bands probe different rest-frame bands.
The same concerns applies for the optical luminosities. Our rest-frame
continuum optical
luminosities are always inferred through rest-frame optical spectra, even at high-z
(through near-IR spectra).
\cite{gallagher07} do not probe directly the optical continuum luminosity of high-z sources
(at high-z they only have optical and Spitzer data, which probe UV and near-IR
rest-frame, respectively). Finally, differences between our and \cite{gallagher07} results
may be simply due to the
different samples. As discussed in the previous section, we avoided
dust reddened targets, thus making us little sensitive to extinction effects, while
\cite{gallagher07} sample may include a larger fraction of reddened objects.

A decreasing dust covering
factor as a function of luminosity must translate into a decreasing
fraction of obscured AGNs as a function of luminosity. The
effect has been noted in various X-ray surveys
\citep{ueda03,steffen03,lafranca05,akylas06,barger05,tozzi06,simpson05},
although the results have  been questioned by other authors
\citep[e.g.][]{dwelly06,treister05,wang07}. The X-ray based studies do not distinguish
between dust and gas and thus probe mostly trends of
the {\it gaseous} absorption. Our result provides an independent
confirmation of these trends. Moreover, our findings are complementary
to those obtained in the X-rays since, instead of the covering factor of gas,
we probe the covering factor of dust.

  In order to compare our findings with the
results obtained from X-ray surveys, we have derived the expected fraction
of obscured AGNs by fitting the dust covering factor versus luminosity
relation with an analytical function. Instead of using the simple power-law
illustrated in Fig.~\ref{fig_miropt} (Eq.~\ref{eq_miropt})
we fit the dependence of the covering factor on luminosity with a broken power-law.
The latter analytical function is preferred both because it provides a statistically
better fit and because a simple power-law would yield a covering factor
larger than unity at low luminosities. As a result we obtain the following best fit 
for the fraction of obscured AGNs as a function of luminosity:
\begin{equation}
\rm f_{obsc} = \frac{1}{1+{{\cal L}_{opt}}^{0.414}}
\label{eq_f2_opt}
\end{equation}
where $\rm f_{obsc}$ is the fraction of obscured AGNs relative to the total and
\begin{equation}
  \rm {{\cal L}_{opt}} = \frac{\lambda
L_{\lambda}(5100\AA )~[erg~s^{-1}]}{10^{45.63}}
\end{equation}
The resulting fit is shown with a dashed line in Fig.~\ref{fig_cf}a.
The fraction of obscured AGNs as a function of luminosity
is also reported with a blue, solid line
in Fig.~\ref{fig_fr_typ2}a.
The shaded area reflects the uncertainty in the
bolometric correction discussed above.

The most recent and most complete
investigation on the fraction of X-ray obscured AGNs as a function of
luminosity has been obtained by Hasinger
\citep[2007, in prep., see also ][]{hasinger04}  who
combined the data from surveys of different areas and limiting fluxes
to get a sample of $\sim$700 objects. We convert from X-ray to optical
luminosity by using the non-linear
relation obtained by \cite{steffen06}. The latter
use flux densities at 2~keV and 2500\AA; we adapt their relation to
our reference optical wavelength (5100\AA) by assuming the optical-UV
spectral slope obtained by \cite{vandenberk01}, and to the 2--10~keV
integrated luminosity (adopted in most X-ray surveys) by assuming
a photon index of --1.7, yielding the relation
\begin{equation}
\rm log [L(2-10~keV)] = 0.721\cdot log[\lambda L_{\lambda}(5100\AA)]+11.78
\label{eq_conv_ox}
\end{equation}
(where luminosities are in units of $\rm erg~s^{-1}$).
Fig.~\ref{fig_fr_typ2}a compares the
fraction of obscured AGN obtained by Hasinger (2007) through X-ray
surveys with our result based on the covering factor of hot dust.
Both have the same trends with luminosity, but the fraction of
obscured AGN expected from the hot dust covering factor is systematically
higher. Such an offset is however expected. Indeed,
current high redshift X-ray surveys  do not probe
the Compton thick population of obscured AGNs since these are
heavily absorbed even in the hard X-rays. In local AGNs,
Compton thick nuclei are about as numerous as Compton
thin ones \citep{risaliti99,guainazzi05,cappi06}.
The fraction of
Compton thick,  high luminosity, high redshift AGNs is still
debated, but their contribution certainly makes the fraction of X-ray
obscured AGNs higher than inferred by Hasinger (2007), who can only account
for Compton thin sources. The ratio between the
dust covering factor curve in Fig.~\ref{fig_fr_typ2}a and the X-ray
data from Hasinger (2007), indicates that the ratio between the total number
of obscured AGNs (including Compton thick ones) and Compton thin ones is about 2
even at high luminosities,
i.e. consistent (within uncertainties) with local, low-luminosity AGNs.

An analogous result on the decreasing fraction of obscured AGNs as a
function of luminosity was obtained  by
\cite{simpson05} who compared the numbers of type 2 and type 1 AGNs
at a given L([OIII]$\lambda$5007). To compare with
\cite{simpson05}, we  used our sample to derive the following analytical
description for the
fraction of obscured AGN as a function of L([OIII]$\lambda$5007):
\begin{equation}
\rm f_{obsc} = \frac{1}{1+{{\cal L}_{[OIII]}}^{0.409}}
\label{eq_f2_oiii}
\end{equation}
where $\rm f_{obsc}$ is the fraction of obscured AGNs relative to the total and
\begin{equation}
\rm {{\cal L}_{[OIII]}} = \frac{L([OIII])~[erg~s^{-1}]}{10^{43.21}}
\end{equation}
The corresponding fit is shown with
a dashed line in Fig.~\ref{fig_cf}b, and the fraction of obscured
AGNs as a function of $\rm L([OIII])$ is also shown with a blue line in
Fig.~\ref{fig_fr_typ2}b.  In the latter figure we also
compare the fraction of obscured AGNs obtained by \cite{simpson05} with
our result based on the covering factor of the hot dust.
There is a good agreement (within uncertainties) between the fraction
of obscured AGNs inferred through the two methods, as expected since both
optical surveys and our method probe the covering factor of dust.
However, we shall also mention that the results
obtained by \cite{simpson05} have been questioned by \cite{haas05}, by
arguing that at high luminosities, the derived L([OIII]$\lambda$5007)
may be affected by a large scale absorber.


   \begin{figure*}[!]
   \centering
   \includegraphics[bb=35 165 520 660,width=18cm]{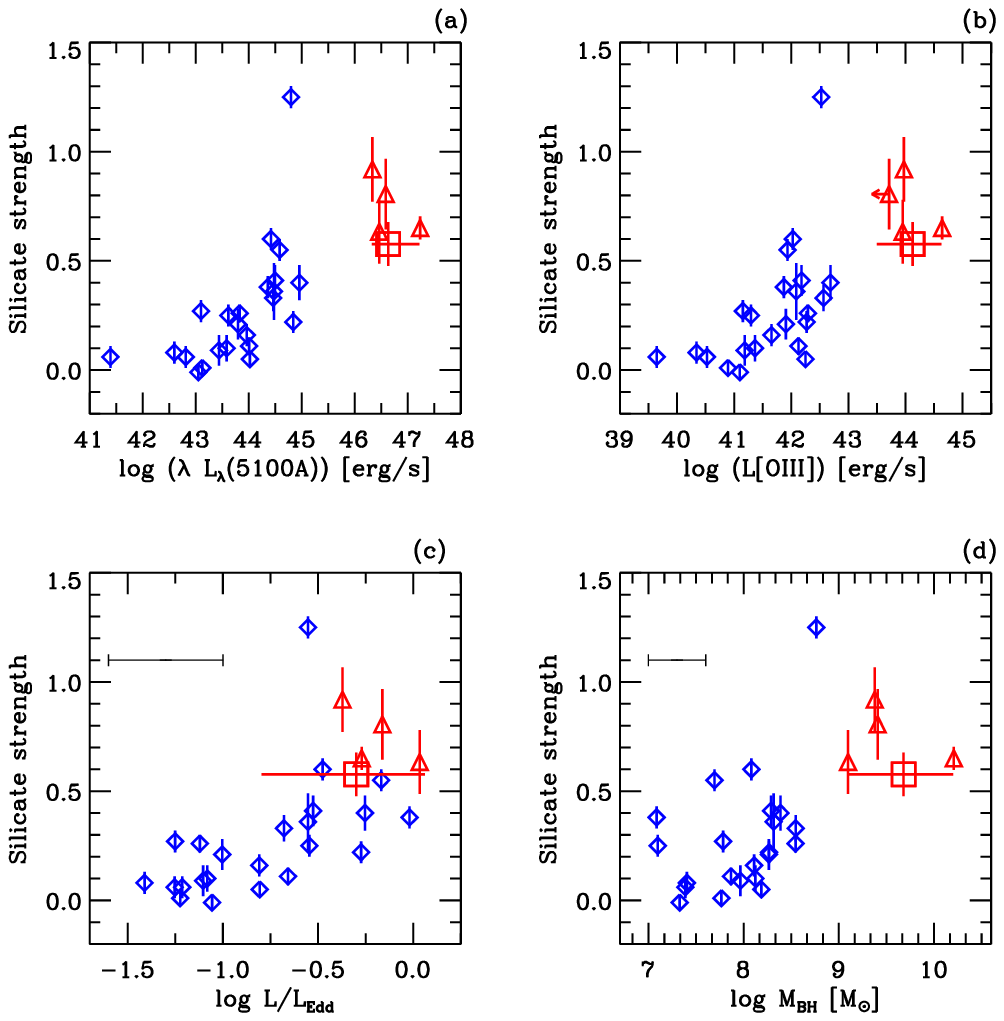}
   \caption{Silicate strength versus (a) \Lop,
     (b) L([OIII]$\lambda$5007), (c) normalized accretion rate ($\rm
     L/L_{Edd}$) and (d) black hole mass.  Symbols are the same as in
     Fig.~\ref{fig_miropt}. The red square indicates the
     silicate strength measured in the average spectrum.
	 The horizontal, black error bars in panels (c) and (d) indicate
	 conservative uncertainties on the accretion rates and BH masses.}
         \label{fig_si}
   \end{figure*}

\subsection{Silicate emission}
\label{sec_si}

In this paper we have presented the most luminous (type 1) QSOs where
Silicate emission has been detected so far. When combined with MIR
spectra of lower luminosity sources, it is possible to investigate
the properties and behavior of this feature over a wide luminosity
range.

The discovery of silicate emission in the spectrum of (mostly type 1)
AGNs obtained by \spitzer\ was regarded as the solution of a long
standing puzzle on the properties of the circumnuclear dusty
medium. Indeed, silicate emission was expected by various models of
the dusty torus. However, the absence of clear detections
 prior to the \spitzer\ epoch induced various authors to
either postulate a very compact and dense torus
\citep[e.g. ][]{pier93} or different dust compositions
\citep{laor93,maiolino01a,maiolino01b}.  Initial \spitzer\
detections of silicate emission relaxed the torus model
assumptions \citep{fritz06}, but more detailed investigations revealed a
complex scenario. The detection of silicate emission even in type
2 AGNs \citep{sturm06b,teplitz06,shi06} suggested that part of the silicate
emission may originate in the Narrow Line Region (NLR)
\citep{efstathiou06}. Further support for a NLR origin of the silicate
emission comes from the temperature inferred for the Silicate
features, which is much lower ($<$200~K) than for the circumnuclear
dust emitting the featureless MIR continuum ($>$500~K), as well as
from MIR high resolution maps spatially resolving the silicate
emission on scales of 100~pc (Schweitzer et al. in prep.).

If most of the observed silicate emission originates in the NLR, then the
effects of circumnuclear hot dust covering factor should be amplified
when looking at the ``silicate strength'' (which we recall is defined
as the ratio of the silicate maximum intensity and the featureless hot
dust continuum).  Indeed, if the covering factor of the circumnuclear
dusty torus decreases, it implies that the MIR hot dust
continuum decreases and the silicate emission increases because a
larger volume of the NLR is illuminated.  Both
effects go in the same direction of increasing the ``silicate
strength''. This scenario is made more complex by the
tendency of the NLR to disappear at very high luminosities, or to get
very dense and not to scale linearly with the nuclear
luminosity \citep{netzer04,netzer06}.
Moreover, the schematic division
of a silicate feature totally emitted by the NLR and a MIR featureless
continuum totally emitted by the inner side of the obscuring torus is
probably too simplistic. There must be at least a small contribution to the
featureless MIR continuum  from dust in the NLR, while
some silicate emission is probably also coming from the obscuring
torus.
However, from a general qualitative point of view we expect a
monotonic behavior of the ``silicate strength'' with the physical
quantity responsible for the changes in the hot dust covering factor.

Before investigating the various trends of the Silicate strength, we
mention that by using the four silicate detections shown in
Fig.~\ref{fig_4si} and listed in Tab.~\ref{tab_ir}, we may in
principle introduce a bias against weak silicate emitters.
 Indeed, although we cannot set useful upper limits on the
silicate strength in most of the other objects, we have likely
missed objects with low silicate strength. However, the mean spectrum
in Fig.~\ref{fig_av} includes all QSOs in our sample, and therefore
its silicate strength should be representative of the average Silicate
emission in the sample (at least for the objects at z$<$2.5, which are
the ones where the observed band includes the silicate feature, and
which are the majority).

Figs.~\ref{fig_si}a-b show the silicate strength of the objects in our
combined sample as  functions of \Lop\ and
 L([OIII]$\lambda$5007). The
red square indicates the silicate strength in the high-z QSO mean
spectrum, while its horizontal bar indicates the range of luminosities
spanned by the subsample of objects at z$<$2.5 (i.e. those
contributing to the silicate feature in the mean spectrum). Low redshift AGNs and
high redshift QSOs show an apparently clear correlation between silicate strength and
luminosity. Although with a significant spread, the Silicate strength is observed
to positively correlate also with the accretion rate $\rm L/L_{Edd}$ and with
the BH mass, as shown in Figs.~\ref{fig_si}c-d. Essentially, the correlations observed
for the Silicate strength reflects the same correlation observed for the
$\rm L(6.7\mu m)/L(5100\AA)$ (with the exception of the accretion rate), in agreement
with the idea that also the Silicate strength is a proxy of the covering factor
of the circumnuclear hot dust, for the reasons discussed above.

Unfortunately, the correlations observed for the Silicate strength do not
improve our understanding on the origin of the decreasing covering factor
with luminosity, i.e. whether the driving physical quantity is the luminosity
itself, the accretion rate or the black hole mass.
Formally, the correlation between Silicate strength and optical continuum
luminosity is tighter than the others (Tab.~\ref{tab_corr}), possibly hinting
at the luminosity itself as the quantity driving the dust covering factor.
However, there are a few low luminosity objects, such as a few LINERs, which
have large silicate strengths \citep{sturm05} and which
clearly deviate from the correlation shown in Fig.~\ref{fig_si}a, thus 
questioning the role of luminosity in determining the Silicate strength.
In addition, the apparently looser correlations of Silicate strength versus accretion
rate and BH mass may simply be due to the additional uncertainties affecting
the latter two quantities (horizontal, black error bars in
Figs.~\ref{fig_si}c,d).

   \begin{figure*}[!]
   \centering
   \includegraphics[bb=15 405 580 670,width=18cm]{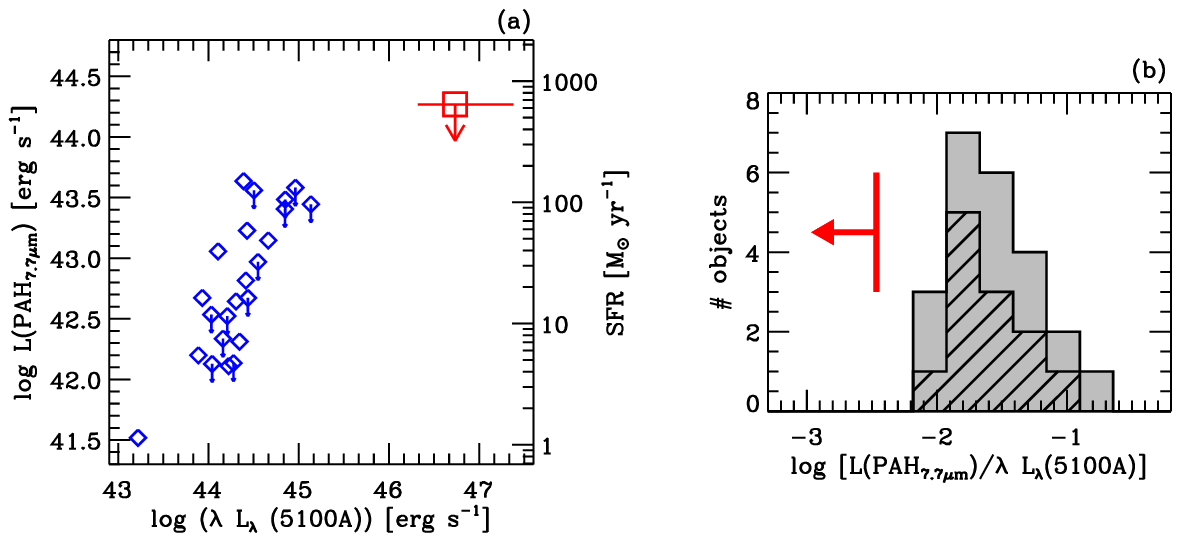}
   \caption{{\it a)} PAH(7.7$\mu$m) luminosity as a function of the
     QSO optical luminosity.  Blue diamonds are data from
     \cite{schweitzer06}. The red square is the upper limit
     obtained by the average spectrum of luminous, high-z QSOs.
	 {\it b)} Distribution of the PAH(7.7$\mu$m)
     to optical luminosity ratio in the local QSOs sample of
     \cite{schweitzer06}. The hatched region indicates upper limits.
     The red vertical line indicate the upper limit inferred from the
     average spectrum of luminous QSOs at high-z.  }
         \label{fig_pah_opt}
   \end{figure*}

\subsection{PAHs and star formation}
\label{sec_pah}

The presence and intensity of star formation in QSOs has been a hotly
debated issue during the past few years. A major
step forward in this debate was achieved by \cite{schweitzer06} through the
\spitzer\-IRS detection of PAH features in a sample of nearby QSOs,
revealing vigorous star formation in these
objects. The analysis also shows that the far-IR emission in these
QSOs is dominated by star formation and that the star forming activity
correlates with the nuclear AGN power. Here we show in
Fig.~\ref{fig_pah_opt} the latter correlation in terms of
PAH(7.7$\mu$m) luminosity versus \Lop\
by using the PAH luminosities from the sample of
\cite{schweitzer06} and the corresponding optical data from
\cite{marziani03}. Although the large fraction of upper limits in the
former sample prevents a careful statistical characterization,
Fig.~\ref{fig_pah_opt} shows a general correlation between QSO optical
luminosity and starburst activity in the host galaxy as traced by the
PAH luminosity.

The scale on the right hand side of Fig.~\ref{fig_pah_opt} translates the 7.7$\mu$m PAH
luminosity into star formation rate (SFR). This was obtained by combining the
average $\rm L(PAH_{7.7\mu m})/L(FIR)$ ratio obtained by
\cite{schweitzer06} for the starburst dominated QSOs in their sample
with the $\rm SFR/L(FIR)$ given in \cite{kennicutt98}, yielding
\begin{equation}
\rm SFR~[M_{\odot}~yr^{-1}] = 3.46~10^{-42}~L(PAH_{7.7\mu m})~[erg~s^{-1}]
\label{eq_pah_sfr}
\end{equation}

The average spectrum of high-z, luminous QSOs in Fig.~\ref{fig_av}
does not show evidence for the presence of PAH features and can only
provide an upper limit on the PAH flux relative to the flux at
6.7$\mu$m (since all spectra were normalized to the latter wavelength
prior to computing the average). However, we can derive an upper limit
on the PAH luminosity by assuming the average distance of the sources
in the sample.  The inferred upper limit on the PAH luminosity is
reported with a red square in Fig.~\ref{fig_pah_opt}a, and it
is clearly below the extrapolation of the $\rm L(PAH_{7.7\mu
  m})-\lambda L_{\lambda}(5100\AA)$ relation found for local,
low-luminosity QSOs. This is shown more clearly in
Fig.~\ref{fig_pah_opt}b which shows the distribution of the ratio
$\rm L(PAH_{7.7\mu m})/\lambda L_{\lambda}(5100\AA)$ for local QSOs
(histogram) and the upper limit inferred from the average
spectrum of high-z, luminous QSOs (red solid line). The corresponding
upper limit on the SFR is $\sim 700~M_{\odot}~yr^{-1}$.

Note that certainly there are luminous, high-z QSOs with larger star formation
rates \cite[e.g.][]{bertoldi03,beelen06,lutz07}. However,
since our sample is not pre-selected in terms of MIR or FIR emission,
our result is not biased in terms of star formation and dust content,
and therefore it is representative of the general high-z, luminous QSO
population.

Our results indicate that the relation between star formation
activity, as traced by the PAH features, and QSO power, as traced by \Lop,
 saturates at high luminosity. This
result is not surprising. Indeed, if high-z, luminous QSOs were
characterized by the same average $\rm L(PAH_{7.7\mu m})/\lambda
L_{\lambda}(5100\AA)$ observed in local QSOs, this would imply huge
star formation rates of $\rm \sim 7000~M_{\odot}~yr^{-1}$ at $\rm
\lambda L_{\lambda}(5100\AA)\sim 10^{47}~erg~s^{-1}$. On the contrary,
the few high-z QSOs detected at submm-mm wavelengths have far-IR
luminosities corresponding to SFR of about $\rm 1000-3000~
M_{\odot}~yr^{-1}$ \citep{omont03}. The majority of high-z QSO ($\sim$70\%)
are undetected at submm-mm wavelengths. The mean mm-submm fluxes of
QSOs in various surveys (including both detections and non-detections)
imply SFRs in the range $\rm
\sim 500-1500~M_{\odot}~yr^{-1}$ \citep{omont03,priddey03},
in fair agreement with our finding,
especially if we consider the uncertainties involved in the two different
observational methods to infer the SFR.

We also note that a similar, independent result was obtained by \cite{haas03},
who found that in a large sample of QSOs the ratio between $\rm L_{FIR}$
(powered by star formation) and $\rm L_B$ decreases at high luminosities.

The finding of a ``saturation'' of the relation between star formation
activity and QSO power may provide an explanation for the evolution of
the relation between BH mass and galaxy mass at high redshift. Indeed,
\cite{peng06} and \cite{mclure06}  found that, for a
given BH mass, QSO hosts at z$\sim$2 are characterized by a stellar
mass lower than expected from the local BH-galaxy mass relation. In
other terms, the BH growth is faster, relative to star formation, in high-z,
luminous QSOs.
Our result supports this scenario
by independently showing that the correlation between star formation
and AGN activity breaks down at high luminosities.

\section{Conclusions}
\label{sec_conc}

We have presented low resolution, mid-IR \spitzer\ spectra of a sample
of 25 luminous QSOs at high redshifts ($\rm 2<z<3.5$). We have combined
our data with \spitzer\ spectra of lower luminosity, type-I AGNs, either
published in the literature or in the \spitzer\ archive. The combined
sample spans five orders of magnitude in luminosity, and allowed us
to investigate the dust properties and star formation rate as a
function of luminosity. The spectroscopic
  information allowed us to disentangle the
  various spectral components contributing to the MIR band (PAH and silicate
  emission) and to sample the continuum at a specific $\rm
  \lambda_{rest}$, in contrast to photometric MIR observations.
  The main results are:
\begin{itemize}

\item The mid-IR continuum luminosity at 6.7$\mu$m correlates with the
  optical continuum luminosity but the correlation is not linear. In
  particular, the ratio $\rm \lambda L_{\lambda}(6.7\mu m)/\lambda
  L_{\lambda}(5100\AA )$ decreases by about a factor of ten as a
  function of luminosity over the luminosity range
  $\rm 10^{42.5}<\lambda L_{\lambda}(5100\AA)<10^{47.5}~erg~s^{-1}$.
  This is interpreted as a reduction of the covering factor
  of the circumnuclear hot dust as a function of luminosity.
  This result is in agreement and provides an independent confirmation of
  the recent findings of a decreasing
  fraction of obscured AGN as a function of luminosity, obtained in
  X-ray and optical surveys. We stress that while X-ray surveys
  probe the covering factor of the {\it gas}, our result provides an independent
  confirmation by probing the covering factor of the {\it dust}.
  We have also shown that the dust covering
  factor, as traced by the $\rm \lambda L_{\lambda}(6.7\mu m)/\lambda
  L_{\lambda}(5100\AA )$ ratio, decreases also as a function of the BH
  mass. Based on these correlations alone it is not possible to
  determine whether the physical quantity primarily driving the
  reduction of the covering factor is the AGN luminosity or the BH
  mass.  

\item The mean spectrum of the luminous, high-z QSOs in our sample
  shows a clear silicate emission at $\rm \lambda _{rest}\sim 10\mu
  m$. Silicate emission is also detected in the individual spectra of
  four high redshift QSOs.  When combined with the spectra of local, lower
  luminosity AGNs we find that the silicate strength (defined as the
  ratio between the maximum of the silicate feature and the
  extrapolated featureless continuum) tend to increase as a function of luminosity.
   The silicate strength correlates positively also with the accretion
   rate and with the BH mass, albeit with a large scatter. 

\item The mean MIR spectrum of the luminous, high-z QSOs in our sample
  does not show evidence for PAH emission. Our sample is not pre-selected
  by the FIR emission
  and therefore it is not biased in terms star formation. As a
  consequence, the upper limit on the PAH emission in the
  total mean spectrum provides a useful, representative upper limit on
  the SFR in luminous QSOs at high redshifts. We find that the ratio
  between PAH luminosity and QSO optical luminosity is significantly
  lower than observed in local, lower luminosity AGNs, implying that
  the correlation between star formation rate and AGN power probably
  saturates at high luminosities. This result may explain the
  evolution of the correlation between BH mass and galaxy stellar mass
  recently observed in luminous QSOs at high redshift.

\end{itemize}


\begin{acknowledgements}
  We are grateful to M.~Salvati for useful comments. We are grateful to
  G.~Hasinger for providing us with some of his results prior to
  publication.
This work is based on observations made with the Spitzer Space Telescope, which is operated by the Jet Propulsion Laboratory, California Institute of Technology under a contract with NASA. Support for this work was provided by NASA under contract 1276513 (O.S.).
  RM acknowledges partial support from the Italian Space Agency (ASI).
  MI is supported by Grants-in-Aid for Scientific Research (16740117).
 OS acknowledges support by the Israel Science Foundation under grant 232/03.

\end{acknowledgements}

\begin{table*}[!h]
\caption{Combined sample of high-z luminous QSO, local QSO and Sy1, and physical properties
inferred from optical-UV spectra.}
\label{tab_opt}
{\centering
\begin{tabular}{lcccccccccc}
\hline\hline                 
Name & RA(J2000) & Dec(J2000) & z & $\rm log (\lambda L_{\lambda}(5100\AA ))$ &
  $\rm log L([OIII])$ & $\rm log M(BH)^d$ & $\rm L/L_{Edd}^d$ & $\rm \alpha _{opt-UV}$ & Ref. \\
     &           &            &   & (erg/s)  &
	 (erg/s)          & $\rm M_{\odot}$ &                 &                        & \\
\hline\hline                        
 \multicolumn{10}{c}{High-z luminous QSOs}\\
\hline                        

    2QZJ002830.4-281706 &  00:12:21.18 &$-$28:36:30.2  & 2.401 &  46.59 &   44.41 & 9.72 & 0.35 & $-$1.42 &1\\
          LBQS0109+0213 &  01:12:16.91 &$+$02:29:47.6  & 2.349 &  46.81 &   44.51 &10.01 & 0.30 & $-$1.75 &1\\
 $[$HB89$]$0123+257$^a$ &  01:26:42.79 &$+$25:59:01.3  & 2.369 &  46.58 &   44.32 & 9.10 & 1.40 & $-$1.65 &1\\
            HS0211+1858 &  02:14:29.70 &$+$19:12:37.0  & 2.470 &  46.63 &   44.38 &10.11 & 0.16 & $-$0.01 &3,10\\
    2QZJ023805.8-274337 &  02:38:05.80 &$-$27:43:37.0  & 2.471 &  46.58 &$<$43.71 & 9.41 & 0.69 & $-$1.59 &1\\
SDSSJ024933.42-083454.4 &  02:49:33.41 &$-$08:34:54.4  & 2.491 &  46.39 &   44.12 & 9.67 & 0.25 & $-$1.36 &1\\
             Q0256-0000 &  02:59:05.64 &$+$00:11:21.9  & 3.377 &  46.99 &   44.55 &10.11 & 0.19 & $-$0.96 &2\\
             Q0302-0019 &  03:04:49.86 &$-$00:08:13.4  & 3.286 &  46.83 &   45.01 &10.11 & 0.30 & $-$1.66 &2\\
     $[$HB89$]$0329-385 &  03:31:06.34 &$-$38:24:04.8  & 2.435 &  46.72 &   44.31 &10.11 & 0.18 & $-$1.79 &1\\
SDSSJ100428.43+001825.6 &  10:04:28.44 &$+$00:18:25.6  & 3.040 &  46.45 &   44.47 & 9.34 & 0.70 & $-$0.70 &3,11\\
             TON618$^a$ &  12:28:24.97 &$+$31:28:37.6  & 2.226 &  47.32 &$<$44.12 &10.81 & 0.14 & $-$1.27 &1\\
     $[$HB89$]$1318-113 &  13:21:09.38 &$-$11:39:31.6  & 2.306 &  46.90 &   44.32 & 9.76 & 0.62 & $-$0.99 &1\\
 $[$HB89$]$1346-036     &  13:48:44.08 &$-$03:53:24.9  & 2.370 &  46.89 &   43.73 & 9.95 & 0.41 & $-$1.26 &1\\
                  UM629 &  14:03:23.39 &$-$00:06:06.9  & 2.460 &  46.57 &   44.41 & 9.17 & 1.16 & $-$1.40 &1\\
              UM632$^b$ &  14:04:45.89 &$-$01:30:21.9  & 2.517 &  46.55 &   44.04 & 9.44 & 0.61 & $-$1.27 &1\\
        SBS1425+606     &  14:26:56.10 &$+$60:25:50.0  & 3.202 &  47.39 &   45.04 & 9.83 & 1.73 & $-$1.45 &1\\
  $[$VCV01$]$J1649+5303 &  16:49:14.90 &$+$53:03:16.0  & 2.260 &  46.70 &   44.19 & 9.99 & 0.24 & $-$0.86 &3,11\\
SDSSJ170102.18+612301.0 &  17:01:02.18 &$+$61:23:01.0  & 2.301 &  46.35 &$<$43.51 & 9.73 & 0.20 & $-$1.48 &1\\
SDSSJ173352.22+540030.5 &  17:33:52.23 &$+$54:00:30.5  & 3.428 &  47.02 &   44.36 & 9.58 & 1.28 & $-$1.51 &1\\
$[$HB89$]$2126-158$^b$  &  21:29:12.17 &$-$15:38:41.0  & 3.282 &  47.27 &   44.66 & 9.73 & 1.60 &    0.72 &1\\
    2QZJ221814.4-300306 &  22:18:14.40 &$-$30:03:06.0  & 2.389 &  46.55 &   43.95 & 9.28 & 0.89 & $-$1.27 &1\\
2QZJ222006.7-280324     &  22:20:06.70 &$-$28:03:23.0  & 2.414 &  47.23 &   44.64 &10.21 & 0.54 & $-$1.28 &1\\
             Q2227-3928 &  22:30:32.95 &$-$39:13:06.8  & 3.438 &  46.95 &$<$44.02 &10.31 & 0.19 & $-$1.25 &2\\
     $[$HB89$]$2254+024 &  22:57:17.56 &$+$02:43:17.5  & 2.083 &  46.46 &   43.95 & 9.10 & 1.08 & $-$1.37 &1\\
    2QZJ234510.3-293155 &  23:45:10.36 &$-$29:31:54.7  & 2.382 &  46.33 &   43.97 & 9.38 & 0.42 & $-$1.26 &1\\

High-z QSO aver. (z$<$2.5)$^c$       &      &        &       &  46.63 &   44.07 & 9.68 & 0.51 &  & \\

\hline\hline                        
 \multicolumn{10}{c}{Local QSOs and Sy1s}\\
\hline
    Mrk335 & 00:06:19.52 & $+$20:12:10.4  & 0.025  & 43.62 & 41.29 & 7.10 & 0.28 & $-$2.00 &4,14\\
    IIIZw2 & 00:10:30.80 & $+$10:58:13.0  & 0.090  & 44.02 & 42.25 & 8.19 & 0.16 & $-$1.56 &4,13\\
PG0050+124 & 00:53:34.94 & $+$12:41:36.2  & 0.058  & 44.36 & 41.87 & 7.09 & 0.96 & $-$0.91 &4,13\\
PG0052+251 & 00:54:52.10 & $+$25:25:38.0  & 0.155  & 44.46 & 42.57 & 8.55 & 0.21 & $-$2.27 &4,13\\
  Fairall9 & 01:23:45.78 & $-$58:48:20.5  & 0.046  & 43.80 & 41.91 & 8.27 & 0.10 & $-$1.99 &4,13\\
     Mrk79 & 07:42:32.79 & $+$49:48:34.7  & 0.022  & 43.58 & 41.37 & 8.12 & 0.08 & $-$0.83 &9,14 \\
PG0804+761 & 08:10:58.60 & $+$76:02:42.0  & 0.100  & 44.42 & 42.03 & 8.08 & 0.33 & $-$2.64 &4,12\\
    Mrk704 & 09:18:26.00 & $+$16:18:19.2  & 0.029  & 43.44 & 41.18 & 7.97 & 0.08 &     --  &4\\
PG0953+414 & 09:56:52.40 & $+$41:15:22.0  & 0.234  & 44.96 & 42.69 & 8.39 & 0.56 & $-$2.12 &4,13\\
   NGC3516 & 11:06:47.49 & $+$72:34:06.8  & 0.009  & 42.81 & 40.52 & 7.39 & 0.06 & $-$1.09 &7,8,14 \\
PG1116+215 & 11:19:08.60 & $+$21:19:18.0  & 0.176  & 44.84 & 42.27 & 8.27 & 0.53 & $-$2.58 &4,13\\
   NGC3783 & 11:39:01.72 & $-$37:44:18.9  & 0.010  & 43.05 & 41.10 & 7.33 & 0.09 & $-$1.46 &4,13\\
PG1151+117 & 11:53:49.27 & $+$11:28:30.4  & 0.176  & 44.48 & 42.09 & 8.31 & 0.28 & $-$2.50 &4,12\\
   NGC4051 & 12:03:09.61 & $+$44:31:52.8  & 0.002  & 41.39 & 39.64 & 5.32 & 0.06 & $-$0.82 &5,6,14\\
PG1211+143 & 12:14:17.70 & $+$14:03:12.6  & 0.085  & 44.58 & 41.94 & 7.69 & 0.68 & $-$1.32 &4,13\\
   NGC4593 & 12:39:39.42 & $-$05:20:39.3  & 0.009  & 42.60 & 40.34 & 7.40 & 0.04 & $-$1.26 &4,14\\
PG1309+355 & 13:12:17.76 & $+$35:15:21.2  & 0.184  & 44.50 & 42.18 & 8.29 & 0.30 & $-$2.45 &4,12\\
PG1351+640 & 13:53:15.80 & $+$63:45:45.4  & 0.087  & 44.80 & 42.52 & 8.76 & 0.28 & $-$0.86 &4,13\\
   IC4329a & 13:49:19.26 & $-$30:18:34.0  & 0.016  & 43.13 & 40.89 & 7.77 & 0.06 &     --  &4\\
   NGC5548 & 14:17:59.53 & $+$25:08:12.4  & 0.017  & 43.10 & 41.15 & 7.78 & 0.06 & $-$1.73 &4,13\\
    Mrk817 & 14:36:22.06 & $+$58:47:39.3  & 0.033  & 43.96 & 41.65 & 8.11 & 0.16 & $-$0.56 &4,14\\
    Mrk509 & 20:44:09.73 & $-$10:43:24.5  & 0.034  & 44.01 & 42.13 & 7.87 & 0.22 & $-$1.69 &4,13\\
    Mrk926 & 23:04:43.47 & $-$08:41:08.6  & 0.047  & 43.83 & 42.29 & 8.55 & 0.08 &     --  &4\\
\hline\hline                        
\end{tabular}
}

The following quantities are reported in each column: column 1, object name; columns 2-3, coordinates
(J2000); column 4, redshift;
column 5,  log of the continuum luminosity $\rm \lambda L_{\lambda}$ (in units of erg/s)
at the rest frame wavelength 5100\AA ; column 6, log of the [OIII]$\lambda$5007 emission line
luminosity (in units of erg/s); column 7, log of the black hole mass (in units of $\rm M_{\odot}$);
column 8, Eddington ratio $\rm L_{bol}/L_{Edd}$; column 9, optical-to-UV (1450\AA --5100\AA) continuum slope
($\rm F_{\lambda}\propto \lambda ^{\alpha_{opt-UV}}$);
column 11: reference for the optical and UV data: 1 - \cite{shemmer04},
\cite{netzer04} and therein references for UV data,
2 - \cite{dietrich02} and therein references for UV data, 3 - Juarez et al
(in prep.), 4 - \cite{marziani03}, 5 - \cite{suganuma06}, 6 - \cite{peterson00},
7 - \cite{wanders93}, 8 - \cite{ho01}, 9 - \cite{peterson98}, 10 - \cite{engels98},
11 - SDSS DR5 archive,
12- \cite{baskin05}, 13 - \cite{evans04}, 14 - \cite{kaspi05}.
\begin{list}{}{}
\item[$^a$] Radio loud QSOs for which the extrapolation of the radio synchrotron emission to the MIR is near
or above the observed value, hence the 6.7$\mu$m flux is probably dominated by synchrotron
emission; these objects will be excluded from statistical analyses.
\item[$^b$] Radio loud QSOs for which the extrapolation of the radio synchrotron emission to the MIR is well
below the observed value, hence the 6.7$\mu$m flux is likely thermal.
\item[$^c$] Optical luminosities, black hole mass and Eddington ratio for the stacked
spectrum refer to the average values of only the objects at z$<$2.5, i.e. those who contribute to
the Silicate feature observed in the stacked spectrum.
\item[$^d$] As discussed in \cite{shemmer04},
§the uncertainties on the BH masses and accretion rate
are no larger than a factor of two.
\end{list}
\end{table*}

\begin{table*}[!h]
\caption{Infrared properties of the combined sample of high-z luminous QSO, local QSO and Sy1}
\label{tab_ir}
{\centering
\begin{tabular}{lccccc}
\hline\hline                 
Name & $\rm F_{MIR}$ & $\alpha _{MIR}$ &
$\rm \frac{\lambda L_{\lambda}(6.7\mu m)}{\lambda L_{\lambda}(5100\AA)}$ &
  Si strength & Ref.\\
      &    (mJy)         &                 &                       &  & \\
\hline\hline                        
 \multicolumn{6}{c}{High-z luminous QSOs}\\
\hline                        

\object{2QZJ002830.4-281706}&    4.8 & -0.59 &  0.75 &               &1\\
        \object{LBQS0109+0213}&    8.3 & -0.91 &  0.75 &               &1\\
\object{$[$HB89$]$0123+257}$^a$ &    4.9 & -1.69 &  0.76 &               &1\\
           \object{HS0211+1858} &    7.3 & -1.36 &  1.09 &               &1\\
   \object{2QZJ023805.8-274337} &    3.7 & -1.04 &  0.62 & 0.81$\pm$0.16 &1\\
\object{SDSSJ024933.42-083454.4} &    2.2 & -1.26 &  0.57 &               &1\\
            \object{Q0256-0000} &    2.7 & -1.34 &  0.30 &               &1\\
            \object{Q0302-0019} &    3.6 & -1.67 &  0.56 &               &1\\
    \object{$[$HB89$]$0329-385} &    5.5 & -1.74 &  0.65 &               &1\\
\object{SDSSJ100428.43+001825.6} &    2.3 & -1.51 &  0.73 &               &1\\
            \object{TON618}$^a$ &   20.1 & -1.16 &  0.51 & 0.04$\pm$0.01 &1\\
    \object{$[$HB89$]$1318-113} &    6.6 & -1.70 &  0.47 &               &1\\
\object{$[$HB89$]$1346-036}     &   12.9 & -1.68 &  0.99 &               &1\\
                 \object{UM629} &    3.8 & -0.68 &  0.65 &               &1\\
             \object{UM632}$^b$ &    2.1 & -1.77 &  0.40 &               &1\\
        \object{BS1425+606}     &   23.9 & -1.56 &  0.96 &               &1\\
\object{$[$VCV01$]$J1649+5303}   &    9.4 & -1.28 &  1.03 &               &1\\
\object{SDSSJ170102.18+612301.0} &    3.3 & -0.66 &  0.83 &               &1\\
\object{SDSSJ173352.22+540030.5} &    2.0 & -2.88 &  0.21 &               &1\\
\object{$[$HB89$]$2126-158}$^b$  &   19.1 & -1.18 &  1.08 &               &1\\
   \object{2QZJ221814.4-300306} &    4.6 & -1.03 &  0.78 &               &1\\
\object{2QZJ222006.7-280324}     &   16.0 & -1.52 &  0.58 & 0.65$\pm$0.05 &1\\
            \object{Q2227-3928} &    3.0 & -1.70 &  0.38 &               &1\\
    \object{$[$HB89$]$2254+024} &    3.6 & -1.12 &  0.60 & 0.63$\pm$0.15 &1\\
   \object{2QZJ234510.3-293155} &    4.2 & -1.33 &  1.16 & 0.92$\pm$0.15 &1\\

High-z QSO aver. (z$<$2.5)$^c$& & -1.57&       & 0.58$\pm$0.10 & \\

\hline\hline                        
 \multicolumn{6}{c}{Local QSOs and Sy1s}\\
\hline
   \object{Mrk335} & 130. & -1.38 &  1.90 & 0.25$\pm$0.06 &3\\
   \object{IIIZw2} &  52. & -1.09 &  4.02 & 0.05$\pm$0.03 &2,3\\
\object{PG0050+124} & 245. & -0.74 &  3.55 & 0.38$\pm$0.05 &2,3\\
\object{PG0052+251} &  28. & -1.35 &  2.40 & 0.33$\pm$0.06 &2,3\\
 \object{Fairall9} & 146. & -0.90 &  4.83 & 0.21$\pm$0.07 &2,3\\
    \object{Mkr79} & 200. & -1.03 &  2.44 & 0.10$\pm$0.06 &3 \\
\object{PG0804+761} &  88. & -1.81 &  3.38 & 0.60$\pm$0.05 &2,3\\
   \object{Mrk704} & 190. & -0.95 &  5.64 & 0.09$\pm$0.06 &3\\
\object{PG0953+414} &  26. & -1.88 &  1.68 & 0.40$\pm$0.08 &2,3\\
  \object{NGC3516} & 210. & -1.05 &  2.40 & 0.06$\pm$0.05 &3\\
\object{PG1116+215} &  66. & -1.71 &  3.08 & 0.22$\pm$0.05 &2,3\\
  \object{NGC3783} & 315. & -1.08 &  2.69 &-0.01$\pm$0.03 &2,3\\
\object{PG1151+117} &  10. & -2.19 &  1.08 & 0.36$\pm$0.13 &2,3\\
  \object{NGC4051} & 230. & -0.43 &  4.71 & 0.06$\pm$0.05 &3\\
\object{PG1211+143} & 100. & -1.26 &  1.89 & 0.55$\pm$0.05 &2,3\\
  \object{NGC4593} & 184. & -1.14 &  3.61 & 0.08$\pm$0.05 &2,3\\
\object{PG1309+355} &  25. & -1.24 &  2.82 & 0.41$\pm$0.07 &2,3\\
\object{PG1351+640} &  53. & -0.89 &  0.64 & 1.25$\pm$0.05 &2,3\\
  \object{IC4329a} & 487. & -0.67 &  9.00 & 0.01$\pm$0.03 &2,3\\
  \object{NGC5548} &  69. & -1.06 &  1.53 & 0.27$\pm$0.05 &2,3\\
   \object{Mrk817} & 140. & -0.65 &  1.61 & 0.16$\pm$0.06 &3\\
   \object{Mrk509} & 179. & -1.29 &  1.98 & 0.11$\pm$0.04 &2,3\\
  \object{Mrk926} &  55. & -1.40 &  1.77 & 0.26$\pm$0.04 &2,3\\
\hline\hline                        
\end{tabular}
}

The following quantities are reported in each column: column 1, object name;
column 2, continuum flux density at the observed wavelength corresponding
to $\rm \lambda _{rest}=6.7\mu m$
(after removing starburst and stellar components, in units of mJy);
column 3, power-law index ($\rm F_{\lambda}\propto \lambda ^{\alpha}$) fitted to the continuum
 in the 5--8$\mu$m range (starburst component--subtracted);
column 4, ratio of the continuum emission
at 5100\AA \ and at 6.7$\mu$m, $\rm \lambda L_{\lambda}(5100\AA ) / \lambda L_{\lambda}(6.7\mu m)$;
column 5, Silicate strength; column 6: reference for the infrared data: 1 - this work (from Spitzer program 20493) ;
2 - \cite{shi06}; 3 - this work (from Spitzer archival data).
\begin{list}{}{}
\item[$^a$] Radio loud QSOs for which the extrapolation of the radio synchrotron emission to the MIR is near
or above the observed value, hence the 6.7$\mu$m flux is probably dominated by synchrotron
emission; these objects will be excluded from statistical analyses.
\item[$^b$] Radio loud QSOs for which the extrapolation of the radio synchrotron emission to the MIR is well
below the observed value, hence the 6.7$\mu$m flux is likely thermal.
\item[$^c$] The ratio $\rm \lambda L_{\lambda}(5100\AA ) / \lambda L_{\lambda}(6.7\mu m)$
is not defined for the stacked spectrum, since all spectra were normalized to the 6.7$\mu$m flux
before stacking. As a consequence, only the Silicate strength (and more generally the continuum shape)
has a physical meaning for the stacked
spectrum. 
\end{list}
\end{table*}

\begin{table*}[!h]
\caption{Spearman-rank coefficients for the correlations in Figs.\ref{fig_cf}, \ref{fig_si}, \ref{fig_alphair}
and \ref{fig_alphauv}.}
\label{tab_corr}
{\centering
\begin{tabular}{c | c c c c c c}
\hline\hline
    & $\rm log(\lambda L_{\lambda}(5100\AA))$ & $\rm log(L[OIII])$ & $\rm log(L/L_{Edd})$ &
		$\rm log(M_{BH})$ & $\alpha _{MIR}$ & $\alpha _{opt-UV}$ \\ 
\hline
$\rm \frac{\lambda L_{\lambda}(6.7\mu m)}{\lambda L_{\lambda}(5100\AA)}$ &
 $\rm -0.76~(<10^{-6})$ &  $\rm -0.72~(<10^{-6})$ &  $\rm -0.44~(3~10^{-4})$ & 
	$\rm -0.70~(<10^{-6})$ & $0.29~(0.05)$ & $-0.20~(0.19)$ \\
$\rm log(Si~str.) $ & $\rm 0.83~(2~10^{-5})$ &  $\rm 0.75~(6~10^{-6})$ &  $\rm 0.75~(8~10^{-6})$ & 
	$\rm 0.66~(8~10^{-4})$ & & \\
\hline
\end{tabular}
}\\

Numbers in parenthesis give the probability for the correlation coefficient to deviate from zero.
\end{table*}

\end{document}